# Analysis of dynamic restricted mean survival time based on pseudo-observations[§]

Zijing Yang[1,2], Chengfeng Zhang[2], Yawen Hou[3*], Zheng Chen[2*]

[1]Stomatological Hospital, Southern Medical University and Guangdong Provincial Stomatological Hospital, 366 Jiangnan Avenue South, Guangzhou, 510260, China

[2]Department of Biostatistics, School of Public Health (Guangdong Provincial Key Laboratory of Tropical Disease Research), Southern Medical University, Guangzhou, 510515, China

[3]Department of Statistics, School of Economics, Jinan University, Guangzhou, 510632, China

[*]**corresponding author**

May 2022

---

SUMMARY: In clinical follow-up studies with a time-to-event end point, the difference in the restricted mean survival time (RMST) is a suitable substitute for the hazard ratio (HR). However, the RMST only measures the survival of patients over a period of time from the baseline and cannot reflect changes in life expectancy over time. Based on the RMST, we study the conditional restricted mean survival time (cRMST) by estimating life expectancy in the future according to the time that patients have survived, reflecting the dynamic survival status of patients during follow-up. In this paper, we introduce the estimation method of cRMST based on pseudo-observations, the construction of test statistics according to the difference in the cRMST (cRMSTd), and the establishment of the robust dynamic prediction model using the landmark method. Simulation studies are employed to evaluate the statistical properties of these methods, which are also applied to two real examples. The simulation results show that the estimation of the cRMST is accurate and the cRMSTd test performs well. In addition, the dynamic RMST model has high accuracy in coefficient estimation and better predictive performance than the "static" RMST model. The hypothesis test proposed in this paper has a wide range of applicability, and the dynamic RMST model can predict patients' life expectancy from any prediction time, considering the time-dependent covariates and time-varying effects of covariates.

# 1. Introduction

Time-to-event endpoints are often used in clinical follow-up studies in which the hazard ratio (HR) has been widely reported to evaluate the treatment effect. However, the HR, which reflects the ratio of hazard rates between treatment groups, does not match exactly with the survival rate commonly used in statistical description (Horiguchi et al., 2018), and it is hard to understand or interpret without the proportional hazards (PH) assumption (Callegaro and Spiessens, 2017; Royston, 2015; Uno et al., 2014). An alternative treatment outcome measure is the restricted mean survival time (RMST) (Dehbi, Royston and Hackshaw, 2017; Hasegawa et al., 2020). The RMST $\mu(\tau)$ of a random time-to-event variable $T$ is the mean of the survival time $X = \min(T, \tau)$ limited to a prespecified time point $\tau$, which can be easily estimated by the area under the Kaplan–Meier survival curve $S(t)$ from $t = 0$ to $t = \tau$ : $\int_0^\tau S(t)dt$ ($= E(X) = \mu(\tau)$). Compared with the HR, the RMST is a robust and clinically interpretable summary measure of the survival time distribution that does not rely on the PH assumption. From a statistical point of view, the RMST is the mean survival time from the start of follow-up to a specific time point $\tau$, and from a practical point of view, it can be viewed as the $\tau$-year life expectancy (Royston and Parmar, 2011).

However, the RMST only calculates the mean survival time within the specific time window $[0, \tau]$, which gives the initial prognosis but does not reflect how the prognosis changes over time. For example, patients who undergo surgery are at high risk of death during treatment due to the possibility of post-operative infection and/or transplant rejection, while the life expectancy is greatly increased once they survive the treatment. The life expectancy for further $w$ time, given that a patient has already survived $s$ time after the start of follow-up, $E(\min(T-s, w) | T > s)$, is defined as the conditional restricted mean survival time (cRMST) (Yang et al., 2021), represented by $\mu(s, w)$, where $s \in [0, \tau - w]$ is the prediction time and $w$ is the prediction window. Compared with the RMST, which is only calculated from the start of follow-up, the cRMST provides valuable and relevant information on the dynamic change of the survival status of patients during follow-up.

When considering the impacts of multiple covariates on a patient's life expectancy, most studies directly model the RMST and predict the life expectancy of patients based on the covariate information at the baseline (Andersen, Hansen and Klein, 2004; Tian, Zhao and Wei, 2014; Wang and Schaubel, 2018; Zhong and Schaubel, 2022). These models are limited, however, because they do not account for the changes in these covariates (Thomas and Reyes, 2014), or they only consider the dichotomous time-varying covariate with, at most, one change from untreated to treated (Zhang *et al*., 2022). To overcome this challenge, Lin et al. (2018) extended the traditional regression model by using functional principal component analysis to extract the dominant features of the biomarker trajectory of each individual as time-dependent covariates to conduct dynamic predictions. But as pointed out by the authors, the proposed model focuses on prediction rather than coefficient interpretation. In contrast, the dynamic prediction model proposed by Yang et al. (2021) performs well in both prediction and interpretation. However, this model only calculates the point estimation of regression coefficients (without associated interval estimation) and underestimates the variance of the coefficients, which directly reduces the validity of the hypothesis tests. On this basis, we extend the dynamic RMST model with an improved algorithm, providing a more robust estimation of the variance and confidence intervals for regression analysis and individual prediction.

The rest of this article is organized as follows. In Section 2, we introduce the methodology for the estimation of cRMST, the construction of test statistics based on the difference in cRMST (cRMSTd) between two groups, and the establishment of the landmark dynamic prediction model. In Section 3, we perform extensive simulation studies to assess the performance of our proposed methods. We also illustrate the application of our methods to two example datasets (univariate and multivariate) in Section 4, and we conclude with a discussion in Section 5.

## 2. Methods

### 2.1 Estimation of cRMST Using Pseudo-observations

For a particular prediction time $s$, an (at least approximately) unbiased estimate of cRMST ($\hat{\mu}_{KM}(s,w)$) is obtained using $E(\min(T-s,w)|T>s) = \int_{s}^{s+w} S(t)dt \Big/ S(s)$ or by calculating the area under the Kaplan–Meier survival curve between [$s$, $s+w$] using only those subjects remaining at risk at time $s$ (denoted as $R_s$; the sample size is $N_s$). It can be verified that both approaches lead to identical results (see Web Appendix A for details).

For each subject $i$ ($i = 1,2,\ldots,N_s$) from $R_s$, repeat this calculation without using subject $i$'s data. Hence, each subject will have an estimate for cRMST that specifically excludes itself, labeled $\hat{\mu}_{KM}^{-i}(s,w)$. The $i^{\text{th}}$ pseudo-observation (Andersen et al., 2004; Andersen, Klein and Rosthøj, 2003; Andersen and Perme, 2010) of $\mu(s,w)$ is defined as in Yang et al. (2021):

$$\hat{\mu}_i(s,w) = N_s \cdot \hat{\mu}_{KM}(s,w) - (N_s - 1)\hat{\mu}_{KM}^{-i}(s,w).$$

The cRMST based on the pseudo-observations can be estimated as:

$$\hat{\mu}(s,w) = \frac{1}{N_s}\sum_{i=1}^{N_s}\hat{\mu}_i(s,w),$$

with the estimated variance term:

$$\widehat{Var}(\hat{\mu}(s,w)) = \frac{1}{N_s(N_s-1)}\sum_{i=1}^{N_s}(\hat{\mu}_i(s,w) - \hat{\mu}(s,w))^2.$$

### 2.2 Hypothesis Test

To compare the difference in dynamic life expectancy between treatment groups, we perform a hypothesis test on the cRMSTd between two groups. For the given prediction time $s$ and prediction window $w$, the hypothesis of interest is:

$$H_0: \Delta = \mu_1(s,w) - \mu_0(s,w) = 0, \quad H_1: \Delta \neq 0,$$

where $\mu_0(s,w)$ and $\mu_1(s,w)$ represent the cRMST of the control and treatment groups, respectively. Let $\hat{\mu}_g(s,w)$ and $\widehat{Var}(\hat{\mu}_g(s,w))$, $g = 0,1$, be the estimated values of cRMST and its variance in each group. Under the null hypothesis, the test statistic can be computed as:

$$Z = \frac{\hat{\mu}_1(s,w) - \hat{\mu}_0(s,w)}{\sqrt{\widehat{Var}(\hat{\mu}_0(s,w)) + \widehat{Var}(\hat{\mu}_1(s,w))}} \sim N(0,1).$$

The $1-\alpha$ confidence interval of the cRMSTd between groups is estimated as $(\hat{\mu}_1(s,w) - \hat{\mu}_0(s,w)) \pm Z_{1-\alpha/2}\sqrt{\widehat{Var}(\hat{\mu}_0(s,w)) + \widehat{Var}(\hat{\mu}_1(s,w))}$, where $Z_{1-\alpha/2}$ is the $(1-\alpha/2)$ 100$^{th}$ quantile of the standard normal distribution.

### 2.3 Dynamic Prediction Model

In general terms, the procedure to obtain dynamic predictions using landmarking (Nicolaie et al., 2013; Van Houwelingen, 2007; Yang et al., 2020) is to select a set of landmark time points $0 \leq s_0 < s_1 < ... < s_l < ... < s_L \leq \tau - w$ in some time interval $[s_0, s_L]$. Then the corresponding landmark datasets can be constructed to establish a dynamic prediction model, realizing the updating of the predicted value at different prediction times $s \in [s_0, s_L]$. Based on the dynamic prediction model proposed by Yang et al. (2021), we improved the algorithm and proposed a more robust dynamic RMST regression model.

#### 2.3.1 Model for Fixed Landmark Time Points

For a random landmark time point $s_l$, the corresponding landmark dataset $R_l$ can be constructed by selecting all subjects at risk at $s_l$ and incorporating the current values of any time-dependent covariates. Let $\mathbf{Z}(s_l)$ represent the $P$-dimensional vector of time-fixed and time-dependent covariates (for the time-dependent covariates, the current value at $s_l$ should be used). After calculating the pseudo-observation of cRMST $\hat{\mu}_i(s_l, w)$ corresponding to each subject $i$ ($i = 1, 2, \ldots, N_l$) in $R_l$, we assume a generalized linear model with:

$$g(\hat{\mu}_i(s_l, w)) = \boldsymbol{\beta}^T(s_l)\mathbf{Z}_i^*(s_l),$$

where $g(\cdot)$ is a link function, $\boldsymbol{\beta}(s_l) = \{\beta_0(s_l), \beta_1(s_l), \ldots, \beta_P(s_l)\}$, $\mathbf{Z}^*(s_l) = \{1, \mathbf{Z}^T(s_l)\}^T$, so that $\beta_0(s_l)$ represents the intercept.

Estimates of the regression parameters $\boldsymbol{\beta}(s_l)$ are based on the generalized estimating equations (GEE) method (Liang and Zeger, 1986):

$$\begin{aligned}\mathbf{U}(\boldsymbol{\beta}(s_l)) &= \sum_{i=1}^{N_l} \mathbf{U}_i(\boldsymbol{\beta}(s_l)) \\ &= \sum_{i=1}^{N_l} \left(\frac{\partial}{\partial \boldsymbol{\beta}(s_l)} g^{-1}(\boldsymbol{\beta}^T(s_l)\mathbf{Z}_i^*(s_l))\right) V_i^{-1} \left(\hat{\mu}_i(s_l, w) - g^{-1}(\boldsymbol{\beta}^T(s_l)\mathbf{Z}_i^*(s_l))\right) = 0\end{aligned},$$

where $g^{-1}(\cdot)$ is the inverse function of $g(\cdot)$, $V_i$ is a working variance of $\hat{\mu}_i(s_l, w)$ (Andersen et al., 2004). The estimators of $\boldsymbol{\beta}(s_l)$ are asymptotically normal $\sqrt{N_l}(\hat{\boldsymbol{\beta}}(s_l) - \boldsymbol{\beta}(s_l)) \sim N(0, \Sigma)$, and the asymptotic variance $\Sigma$ can be consistently estimated (Andersen et al., 2003):

$$\hat{\Sigma} \approx \mathbf{I}(\hat{\boldsymbol{\beta}}(s_l))^{-1} \widehat{Var}(\mathbf{U}(\boldsymbol{\beta}(s_l))) \mathbf{I}(\hat{\boldsymbol{\beta}}(s_l))^{-1},$$

where

$$\mathbf{I}(\hat{\boldsymbol{\beta}}(s_l)) = \frac{1}{N_l} \sum_{i=1}^{N_l} \left(\frac{\partial}{\partial \boldsymbol{\beta}(s_l)} g^{-1}(\hat{\boldsymbol{\beta}}^T(s_l)\mathbf{Z}_i^*(s_l))\right)^T V_i^{-1} \left(\frac{\partial}{\partial \boldsymbol{\beta}(s_l)} g^{-1}(\hat{\boldsymbol{\beta}}^T(s_l)\mathbf{Z}_i^*(s_l))\right),$$

$$\widehat{Var}(\mathbf{U}(\boldsymbol{\beta}(s_l))) = \frac{1}{N_l} \sum_{i=1}^{N_l} \mathbf{U}_i(\hat{\boldsymbol{\beta}}(s_l)) \{\mathbf{U}_i(\hat{\boldsymbol{\beta}}(s_l))\}^T.$$

*2.3.2 Landmark Super Model (Dynamic RMST Model)*

The above approach can be used to fit a regression model at the specific time point $s_l \in [s_0, s_L]$, thereby predicting the survival of patients who had survived to $s_l$ in the future $w$ time. However, this method requires a separate regression model to be fitted at each time point $s_l$, for which the prediction is required. This is not very practical, and some form of smoothing and simplification is needed. To this end, we establish the landmark super model.

After selecting a set of landmark time points $s_j$, $j = 0, 1, ..., L$, the corresponding landmark datasets $R_j$ can be stacked into a "super prediction dataset" $R$, in which the sample size corresponding to each individual $i, i = 1, 2, ..., N$ in the total population is denoted as $n_i$, and the vector $\hat{\mu}_i(s, w) = \{\hat{\mu}_i(s_j, w), j \in \{0, 1, ..., n_i - 1\}\}$ is the dynamic pseudo-observations for the $i^{th}$ individual at different time points $s_j$. Given a link function $g(\cdot)$, the generalized linear model is established based on the dataset $R$:

$$g(\hat{\mu}_i(s, w)) = \boldsymbol{\beta}^T(s)\mathbf{Z}_i^*(s),$$

where $\boldsymbol{\beta}(s) = \{\beta_0(s), \beta_1(s), ..., \beta_P(s),\}$, $\mathbf{Z}^*(s) = \{1, \mathbf{Z}^T(s)\}^T$.

To model the time-dependent behavior of $\boldsymbol{\beta}(s)$ across $s \in [s_0, s_L]$, we can employ a linear model for the $p^{th}$ component of $\boldsymbol{\beta}(s)$:

$$\beta_p(s) = \boldsymbol{\beta}_p^T \mathbf{h}_p(s), \quad s \in [s_0, s_L],$$

where $\mathbf{h}_p(s)$ is a suitable set of basis functions, and $\boldsymbol{\beta}_p (p = 0, 1, ..., P)$ is a vector of coefficients (Nicolaie et al., 2013). Different covariates can use different basis functions. Defining $\boldsymbol{\beta} = \{\boldsymbol{\beta}_0, \boldsymbol{\beta}_1, ..., \boldsymbol{\beta}_P\}$ to be a vector of length $q$, the vector $\boldsymbol{\beta}(s)$ can be written as $\mathbf{H}(s)\boldsymbol{\beta}$, with $\mathbf{H}(s)$ a $(P+1) \times q$ matrix containing the basis functions. The estimating equations to be solved are:

$$\mathbf{U}(\boldsymbol{\beta}) = \sum_{i=1}^{N} \left( \frac{\partial}{\partial \boldsymbol{\beta}} g^{-1}((\mathbf{H}\boldsymbol{\beta})^T \mathbf{Z}_i^*) \right)^T \mathbf{V}_i^{-1} (\hat{\mu}_i(s, w) - g^{-1}((\mathbf{H}\boldsymbol{\beta})^T \mathbf{Z}_i^*)) = 0,$$

where $g^{-1}((\mathbf{H}\boldsymbol{\beta})^T \mathbf{Z}_i^*) = \{g^{-1}((\mathbf{H}(s_0)\boldsymbol{\beta})^T \mathbf{Z}_i^*(s_0)), ..., g^{-1}((\mathbf{H}(s_{n_i-1})\boldsymbol{\beta})^T \mathbf{Z}_i^*(s_{n_i-1}))\}$ is an $n_i$-dimensional vector, and $\mathbf{V}_i$ is a working covariance matrix of $\hat{\mu}_i(s, w)$ with a predefined structure (Nicolaie et al., 2013; Zhao et al., 2020), which reflects the correlation between the pseudo-observations of the same individual at different landmark time points. This paper considers the most commonly used independent working covariance matrix.

A sandwich estimator is used to estimate the asymptotic variance–covariance matrix of $\hat{\boldsymbol{\beta}}$:

$$\hat{\boldsymbol{\Sigma}} \approx \mathbf{I}(\hat{\boldsymbol{\beta}})^{-1} \widehat{Var}(\mathbf{U}(\boldsymbol{\beta})) \mathbf{I}(\hat{\boldsymbol{\beta}})^{-1},$$

where

$$\mathbf{I}(\hat{\boldsymbol{\beta}}) = \frac{1}{N} \sum_{i=1}^{N} \left( \frac{\partial}{\partial \boldsymbol{\beta}} g^{-1}((\mathbf{H}\hat{\boldsymbol{\beta}})^T \mathbf{Z}_i^*) \right)^T \mathbf{V}_i^{-1} \left( \frac{\partial}{\partial \boldsymbol{\beta}} g^{-1}((\mathbf{H}\hat{\boldsymbol{\beta}})^T \mathbf{Z}_i^*) \right),$$

$$\widehat{Var}(\mathbf{U}(\boldsymbol{\beta})) = \frac{1}{N} \sum_{i=1}^{N} \mathbf{U}_i(\hat{\boldsymbol{\beta}}) \mathbf{U}_i(\hat{\boldsymbol{\beta}})^T.$$

When considering the independent covariance structure—that is, assuming that the pseudo-observations corresponding to the same individual are independent of each other—the calculation process of the information matrix can be transformed to directly calculate each observation in the super prediction dataset $R$:

$$\mathbf{I}(\hat{\boldsymbol{\beta}}) = \frac{1}{N} \sum_{i=1}^{N} \sum_{k=1}^{n_i} \left( \frac{\partial}{\partial \boldsymbol{\beta}} g^{-1}((\mathbf{H}\hat{\boldsymbol{\beta}})^T \mathbf{Z}_i^*(s_{k-1})) \right)^T V_{ik}^{-1} \left( \frac{\partial}{\partial \boldsymbol{\beta}} g^{-1}((\mathbf{H}\hat{\boldsymbol{\beta}})^T \mathbf{Z}_i^*(s_{k-1})) \right),$$

where $V_{ik}$ is the variance of the pseudo-observations $\hat{\mu}_i(s_{k-1}, w)$ of individual $i$ at $s_{k-1}$. However, it is worth noting that the denominator $N$ reflects the number of individuals rather than the total number of pseudo-observations of these individuals at each landmark time point (that is, the number of rows in $R$). The same is true for $\widehat{Var}(\mathbf{U}(\boldsymbol{\beta}))$. Incorrect use will result in an underestimate of $\boldsymbol{\Sigma}$.

Let $\tilde{\mathbf{Z}}(s)$ be the covariate vector for a new patient at any prediction time $s \in [s_0, s_L]$; the cRMST in the future $w$ time can be predicted by:

$$\hat{\mu}(s, w) = g^{-1}(\hat{\boldsymbol{\beta}}(s)^T \tilde{\mathbf{Z}}^*(s)) = g^{-1}((\mathbf{H}(s)\hat{\boldsymbol{\beta}})^T \tilde{\mathbf{Z}}^*(s)),$$

where $\tilde{\mathbf{Z}}^*(s) = (1, \tilde{\mathbf{Z}}(s)^T)^T$. The corresponding $1-\alpha$ two-sided confidence interval is estimated by $\hat{\mu}(s,w) \pm t_{\alpha/2,v} S_{\hat{\mu}(s,w)}$, where $v$ represents the degrees of freedom, and the standard error of $\hat{\mu}(s,w)$ is estimated consistently by:

$$S_{\hat{\mu}(s,w)} = \sqrt{\widehat{Var}(g^{-1}(\hat{\boldsymbol{\beta}}(s)^T \boldsymbol{Z}^*(s)))}$$
$$= \sqrt{\left(\frac{d}{d\hat{\boldsymbol{\beta}}} g^{-1}(\hat{\boldsymbol{\beta}}(s)^T \boldsymbol{Z}^*(s))\right)^T \widehat{Var}(\hat{\boldsymbol{\beta}}) \left(\frac{d}{d\hat{\boldsymbol{\beta}}} g^{-1}(\hat{\boldsymbol{\beta}}(s)^T \boldsymbol{Z}^*(s))\right)},$$

where

$$\frac{d}{d\hat{\boldsymbol{\beta}}} g^{-1}(\hat{\boldsymbol{\beta}}(s)^T \boldsymbol{Z}^*(s)) = \left(\frac{dg^{-1}(x)}{dx}\right)_{|x=\hat{\boldsymbol{\beta}}(s)^T \boldsymbol{Z}^*(s)} \frac{\partial}{\partial \hat{\boldsymbol{\beta}}}(\hat{\boldsymbol{\beta}}(s)^T \boldsymbol{Z}^*(s))$$
$$= \left(\frac{dg^{-1}(x)}{dx}\right)_{|x=\hat{\boldsymbol{\beta}}(s)^T \boldsymbol{Z}^*(s)} \mathbf{H}(s)^T \boldsymbol{Z}^*(s).$$

## 3. Simulation Studies

### 3.1 *Parameter Estimation and Hypothesis Test of the cRMSTd*

#### 3.1.1 *Simulation Setup*

A Monte Carlo simulation study was designed to explore the accuracy of estimation of the cRMST and to evaluate the performance of the test method. Four simulation scenarios were formulated under both the PH assumption and the non-PH assumption to render a wide spectrum of potential scenarios that can be encountered in follow-up studies (Web Figure 1): 1) no difference between groups; 2) a PH assumption with HR = 0.67; 3) an early difference between groups; and 4) a late difference with curves separated at 10 months. The survival time $T$ of the control group was generated based on an exponential distribution with a median of 10 months. In the simulation scenarios under the non-PH assumption, the HR was a piecewise function: in Scenario 3, HR was equal to 0.1 up to 5 months, 0.67 from 5 to 15 months, and 1.0 after 15 months; in Scenario 4, HR was equal to 1.0 up to 10 months and 0.33 after 10 months.

The prediction time $s$ was fixed at 5 and 10 months, and the prediction window $w$ was set as 5, 10, and 15 months. The censoring time $C$ in the two groups was generated from uniform distributions $U(0, a)$ and $U(0, b)$, respectively. Then, each individual was assigned an observed time $Y = \min(T, C)$ and the event indicator $\delta = I(T < C)$. The censoring rate in each group was controlled at approximately 0%, 15%, and 30% by changing the censoring parameters $a$ and $b$. Sample size scenarios were 100, 200, and 500 per group. All simulations were performed using $n_{sim} = 10,000$ iterations, and the significance level was fixed at 0.05.

To mimic the population, we generated survival time $T$ of $N = 1,000,000$ subjects according to the distribution of each group, and calculated the "true" cRMST by averaging the restricted survival time $X = \min(T, s+w)$ of these subjects who were still at risk at $s$. Then, according to the above settings, we sampled from these large populations and measured the performance of the cRMSTd estimation by calculating the following indicators (Morris, White and Crowther, 2019):

1) bias: the difference between the average of the 10,000 estimated values and the "true" value $\Delta$, $\frac{1}{n_{sim}} \sum_{i=1}^{n_{sim}} \hat{\Delta} - \Delta$;

2) relative bias (Rel bias): $(\frac{1}{n_{sim}} \sum_{i=1}^{n_{sim}} \hat{\Delta} - \Delta) / \Delta$;

3) root mean squared error (RMSE): $\sqrt{\frac{1}{n_{sim}} \sum_{i=1}^{n_{sim}} (\hat{\Delta} - \Delta)^2}$;

4) relative standard error (Rel SE): the ratio of the empirical standard error ($\sqrt{\frac{1}{n_{sim}-1} \sum_{i=1}^{n_{sim}} (\hat{\Delta} - \bar{\Delta})^2}$, where $\bar{\Delta} = \frac{1}{n_{sim}} \sum_{i=1}^{n_{sim}} \hat{\Delta}$) to the average model standard error ($\sqrt{\frac{1}{n_{sim}} \sum_{i=1}^{n_{sim}} [\widehat{Var}(\hat{\Delta})]}$); and

5) empirical coverage probability (CP): the proportion of samples in which the 95% confidence interval of the cRMSTd included $\Delta$.

*3.1.2 Simulation Results*

Table 1 shows the performance of cRMSTd estimation between groups under different scenarios. Considering that the true cRMSTd in Scenario 1 is approximately equal to 0, we used bias rather than relative bias to assess the difference between the estimated values and the true value. In summary, the estimation of cRMSTd has a small bias (or relative bias) under all scenarios (the absolute value of the bias is less than 0.013, and the absolute value of the relative bias is less than 0.023), and the root mean squared error decreases with increasing sample size and decreasing censoring rate. Meanwhile, the relative standard error is approximately equal to 1, and most of the CP falls within a reasonable range, indicating that the accuracy of the estimation method is high.

For the hypothesis tests, under the combination of different sample sizes, censoring rates, and values of $s$ and $w$, the type I error rates fluctuate around 0.05, indicating that the proposed method can well control the type I error. In the power results, it is worth noting that the selection of $s$ and $w$ determines the interval range ($[s, s+w]$) for calculating the mean survival time of patients, which leads to the different results of the same simulated situation under different values of $s$ and $w$. In Scenario 2, for a given prediction time point $s$, the test becomes more powerful with increasing length of the prediction window $w$; conversely, for a given $w$, the power declines with increasing $s$. The results of Scenario 3 show the same trend as Scenario 2. However, in the later follow-up period, the difference between two survival curves in Scenario 3 is smaller than that in Scenario 2, which results in a larger reduction in powers with increasing $s$. In Scenario 4, since the two survival curves do not separate until after 10 months, the power results obtained when $s$ and $w$ are both 5 actually correspond to the type I error rates, which always fluctuate around 0.05. In addition, this method becomes much more powerful with increasing values of $s$ and $w$.

## 3.2 Dynamic Prediction

In this paper, two simulation studies were conducted for the dynamic prediction model: one to verify the accuracy of the regression coefficient estimation in the dynamic RMST model, and the other to evaluate the prediction performance of this model. Both simulation studies generated data based on a joint model of longitudinal data and survival data (Huang et al., 2016; Lin et al., 2018) with the following functional form:

$$\begin{cases} Y_i(t) = m_i(t) + \varepsilon_i(t) \\ h_i(t) = h_0(t)\exp(\gamma_1 X_{1i} + \gamma_2 X_{2i} + \alpha m_i(t)) \end{cases}.$$

In the longitudinal sub-model, the observed values of longitudinal biomarker data of individual $i$ at time $t$ $Y_i(t)$ include the true values $m_i(t)$ and noisy measurement errors $\varepsilon_i(t)$, where $\varepsilon_i(t) \sim N(0, 0.5)$. For $m_i(t)$, the linear mixed-effects model and quadratic mixed-effects model were used to generate longitudinal measures with the following formulas:

$$m_i(t) = (\beta_{01} + b_{0i}) + (\beta_{10} + b_{1i})t + \beta_1 X_{1i} + \beta_2 X_{2i},$$

$$m_i(t) = (\beta_{02} + b_{0i}) + (\beta_{10} + b_{1i})t + (\beta_{20} + b_{2i})t^2 + \beta_1 X_{1i} + \beta_2 X_{2i},$$

where $\boldsymbol{\beta} = \{\beta_{01}, \beta_{02}, \beta_{10}, \beta_{20}, \beta_1, \beta_2\} = \{3, 0.5, -0.2, 0.1, 1, -1\}$ were the fixed effects and $X_i$ were the baseline covariates, including one binary variable $X_{1i}$ and one continuous variable $X_{2i} \sim N(1,1)$. In the linear mixed-effects model, the random effects term was $b_i = (b_{0i}, b_{1i})' \sim N(0, D)$ with $D = \begin{pmatrix} 1 & 0.5 \\ 0.5 & 0.04 \end{pmatrix}$, while in the quadratic mixed-effects model, $b_i = (b_{0i}, b_{1i}, b_{2i})' \sim N(0, D)$, with $D = \begin{pmatrix} 1 & 0.1 & 0.1 \\ 0.1 & 0.36 & 0.1 \\ 0.1 & 0.1 & 0.0025 \end{pmatrix}$. In the survival sub-model, $h_0(t) = \lambda t^{\lambda-1} \exp(\eta)$ was the Weibull baseline hazard function, where the shape parameter $\lambda = 3$ and the scale parameter $\eta = -6$. The remaining coefficients were set as $\gamma_1 = 1$, $\gamma_2 = -1$, and $\alpha = 1$. Each patient was observed once at time 0, and the rest of the observation time was uniformly distributed from 0 to the maximum follow-up time. The maximum number of observations did not exceed 10, and the longest follow-up time did not exceed 20.

*3.2.1 Coefficient Estimation*

Since the true model of the dynamic RMST model was unknown, the "true" values of the regression coefficient were determined in the following ways:

1) Generated $N = 1,000,000$ subjects as the total population;

2) Defined the prediction window $w = 5$ and prediction time $s \in [s_0, s_L] = [0,10]$. The landmark time points were selected as $s_j = 0, 0.5, 1, ..., 10$, and the landmark datasets $R_j, j = 0, 1, ..., 20$ can be established correspondingly;

3) Stacked all the landmark datasets into a super prediction dataset $R$, and a dynamic RMST model can be fitted to obtain the "true" values of the regression coefficients, where $g(\cdot)$ is the identity link function and $\mathbf{h}_p(s)$ is the natural cubic spline function, with knots at 2, 4, 6, and 8. In addition, for numeric stability, the prediction time was standardized using $\bar{s} = s/(s_L - s_0) = s/10$.

Either 500 or 1,000 subjects were randomly selected from the total population, with the censoring rate remaining around 0%, 15%, and 30% by changing the distribution parameter $a$ of the censoring time $C \sim U(0,a)$. The same dynamic RMST model was fitted to these subjects, and the bias, relative bias, root mean squared error, relative standard error, and empirical CP of each regression coefficient compared with its "true" value were calculated through 10,000 simulations.

The results shown in Table 2 suggest that the regression coefficients are well estimated with very small bias and good coverage. Additionally, the relative standard errors are close to 1, and the root mean squared errors are also reasonably low, also proving good model performance.

*3.2.2 Predictive Performance*

We validated the dynamic predictions by evaluating both Harrell's C-index (Harrell, Lee and Mark, 1996) and prediction error (Tian et al., 2007) (that is, the difference between the predicted cRMST and its true value). In each scenario, separate training and validation data sets were generated, with $n = 500$ (or 1,000) subjects in the training set and $n = 300$ subjects in the validation set. The censoring rate was also controlled at approximately 0%, 15%, and 30%. The training data set was used to build a dynamic RMST model, and the cRMSTs were predicted for subjects still alive at $s_j$ in the validation data set. Then the C-index and prediction error were calculated at each landmark time point $s_j, j = 0,1,...,20$. At the same time, the RMST model (Andersen et al., 2004) with $\tau = s_j + w$ was also established for comparison with the dynamic RMST model. The performance of these two models in the validation data was summarized across 1,000 simulated iterations, and the average C-index and prediction error values were calculated.

As shown in Figure 1, the results obtained with different sample sizes and censoring rates are basically consistent. Compared with the RMST model, the dynamic RMST model presents a higher C-index and a lower prediction error with increasing prediction time *s*. Combining the results of these two indicators, it can be concluded that the dynamic RMST model proposed in this paper has very high predictive performance.

## 4. Illustrative Examples

*4.1 Univariate Analysis (Hypothesis Test)*

In a study evaluating the effects of thoracic radiotherapy for extensive-stage small-cell lung cancer (Slotman et al., 2015), a total of 495 enrolled patients were randomly assigned (1:1 randomization ratio) to receive either thoracic radiotherapy plus prophylactic cranial irradiation (experimental group) or prophylactic cranial irradiation only (control group). The primary endpoint was overall survival. Figure 2A shows that the survival curves for the two groups started to diverge after about 8 months, which indicates the possible presence of a delayed treatment effect. Meanwhile, the Grambsch-Therneau test (Grambsch and Therneau, 1994) showed that there is clear evidence of non-PH ($\chi^2$ = 10.200, $P$ = 0.001), so the power of the log-rank test ($\chi^2$ = 3.497, $P$ = 0.061) was low. In addition, because the RMST test (Royston and Parmar, 2013) was not sensitive to this delayed effect (Eaton, Therneau and Le-Rademacher, 2020), it also failed to conclude that there was any difference between the two groups ($Z$ = 1.705, $P$ = 0.088) when $\tau = 26.38$ months (the maximum observed event time of each group).

In contrast, the cRMST test can compare the life expectancy of patients in the next $w$ time given those who have already survived $s$ time. A series of prediction time points $s$ were selected, and the differences in 12-month ($w$ = 12) cRMST between groups corresponding to each time point $s$ (red solid line in Figure 2B) and its 95% confidence interval (red dashed lines in Figure 2B) were calculated. Web Table 1 shows the results of the cRMSTd test at several specific prediction time points. The results show that in the early follow-up period (i.e., the prediction time $s$ less than 6 months), there was no statistical difference in the 12-month life expectancy between the two groups of patients who were still alive at $s$ (the 95% confidence interval included 0). After 6 months, for patients who had already survived for $s$ time from the start of follow-up, receiving thoracic radiotherapy plus prophylactic cranial irradiation improved their 12-month life expectancy more than prophylactic cranial irradiation alone. To facilitate comparison with the RMSTd test, the ($s$+12)-month ($\tau = s + 12$) RMST differences between the two groups were also calculated (green lines in Figure 2B). The results of the RMST test indicate that there are no statistical differences between groups.

### 4.2 *Multivariate Analysis (Dynamic Prognosis and Prediction)*

The data of 407 patients with chronic kidney disease who received renal transplantations in the hospital of the Catholic University of Leuven (Belgium) between 1983 and 2000 (Rizopoulos and Ghosh, 2011) were selected. The time elapsed from renal transplantation to graft failure was the primary outcome. From the 407 patients, 126 suffered a graft failure; this corresponds to a 69% censoring. During the study follow-up, multiple biomarkers, including the blood hematocrit level, the urinary protein content (proteinuria), and the glomerular filtration rate (GFR), were measured periodically to test kidney conditions. In addition, there were three baseline covariates, including the patient's age, sex, and weight.

To obtain the dynamic prediction of 5-year ($w=5$) cRMST, landmark time points $\{s_j, j=0,1,...,20\}=\{s_0, s_1,...,s_{20}\}$ were chosen every 6 months between 0 ($s_0 = 0$) and 10 ($s_L = 10$) years after renal transplantations. At each time point $s_j$, the corresponding landmark dataset was established to fit the dynamic RMST model with the identity link function. In this model, $\mathbf{h}_p(s)$ was a natural cubic spline function with five degrees of freedom, and the prediction time was standardized using $\bar{s} = s/(s_L - s_0)$. A backward stepwise model selection procedure was used based on the quasi-likelihood under the independence model criterion (QIC) (Pan, 2001).

The results of the dynamic RMST model are shown in Web Table 2. The data show that all covariates except sex ($Z = -1.901$, $P = 0.057$) have statistically significant effects on graft life expectancy. Among them, three longitudinal covariates demonstrated significant time-varying effects on the 5-year cRMST, which is shown in Figure 3. The higher the levels of the hematocrit and GFR, the higher the life expectancy of graft, while the effect of the proteinuria on life expectancy is the opposite. In addition, the effect of hematocrit and proteinuria increased with increasing prediction time, while GFR showed a trend of first increasing and then decreasing. A "static" RMST regression model (Andersen et al., 2004) with $\tau = 15$ years was also established (Web Table 3) for comparison with the dynamic RMST model, which showed that only hematocrit was statistically significant.

In addition to exploring the dynamic effects of covariates on graft life expectancy, the dynamic RMST model can also provide individual dynamic predictions for patients. Three example patients (see Web Table 4 for details) were selected from the dataset, and the 5-year cRMSTs (with 95% confidence intervals) of graft corresponding to these patients predicted by the dynamic RMST model at different prediction times are shown in Figure 4 (black lines). For patient A, in the early follow-up period, the 5-year life expectancy of graft was basically unchanged, indicating that the condition remained stable. However, at about 7.59 years after kidney transplantation, the clinical situation suddenly worsened, the proteinuria increased, and GFR rapidly decreased, resulting in a significant reduction in graft life expectancy (Figure 4A). In contrast, patient B's condition continued to deteriorate, showing the decrease in graft life expectancy continued after transplantation (Figure 4B). Similar to patient A, at about 5.14 years after transplantation, patient C had an increase in the proteinuria and a decrease in GFR, accompanied by a decrease in hematocrit, resulting in a greatly reduced graft life expectancy. But the situation improved around 6 years after transplantation. Although the GFR still showed a downward trend, the proteinuria returned to normal, and the graft life expectancy increased at this time (Figure 4C). For comparison, the prediction results of the RMST model are also presented in Figure 4 (gray lines; e.g., $s = 0$ corresponds to the 5-year RMST from the start of follow-up, and $s = 10$ corresponds to the 15-year RMST from the start of follow-up). Since only the information at the start of follow-up ($s = 0$) was considered, the RMST of these patients always maintained an upward trend, which was only a cumulative process and did not reflect the change in life expectancy over prediction time.

After the original dataset was divided into a training set and a validation set with the ratio of 7:3, cross-validation was performed to calculate the C-index and prediction error separately for each landmark time point $s_j$. This process was repeated 200 times to obtain the average C-index and prediction error values (solid lines in Figure 5). Correspondingly, at each $s_j$, the performance measures of the RMST model ($\tau = s_j + w$) were also calculated (dashed lines in Figure 5). The predictive performance of the dynamic RMST model is significantly better than that of the RMST model, mainly because the RMST model only utilizes the information of the covariates at the baseline, while the dynamic RMST model takes into account the changing information of the covariates during the follow-up period, which can update the predicted value over time.

## 5. Discussion

The RMST only calculates the patient's life expectancy within $[0, \tau]$, which does not reflect how the prognosis changes over time. In view of this, some researchers (Liao, Liu and Wu, 2020; Zhao et al., 2016) suggested constructing an RMST($t^*$) ($t^* \in [0, \tau]$) curve based on the RMST over time to examine the RMST process over the entire time span of interest. Correspondingly, in the example analysis in this paper, the RMST($t^*$) ($t^* = s + w$) difference curve between the two groups was plotted (green lines in Figure 2B), reflecting the change in the treatment effect. Nevertheless, all the values on this curve are calculated from the start of follow-up, but they are still not very informative for a patient (and their treating physician) who has already survived a number of years. As time progresses, the cRMST studied in this paper provides more relevant prognostic information for survivors by estimating life expectancy based on the time patients have survived. For patients, the cRMST is an easily understandable concept and can be used to clearly portray their changing survival profile. Clinicians can also make use of cRMST to understand the patient's condition in real time, effectively guide the clinical treatment, and adjust the treatment plan in time.

Based on the cRMST, we propose an estimation method using pseudo-observations and a hypothesis test method for the difference between two groups. The simulation results (Table 1) show that under the combination of different sample sizes, censoring rates, and values of prediction time *s* and prediction window *w*, the estimation of the cRMSTd is accurate, and the cRMSTd test can well control type I error rates. It is worth noting that the use of cRMST requires a long-term accumulation of information from a large sample, which is mainly due to the fact that cRMST is calculated based on the landmark dataset $R_s$. Because of censoring and early events, $R_s$ is only a subset of the original population. Only when the sample size of the original population is large enough and the follow-up is complete enough can $R_s$ have enough samples for statistical inference. This is why when the PH assumption is satisfied (Scenario 2 in Table 1), the later the prediction time is, the lower the statistical power will be.

However, this long period causes additional problems, as patients' characteristics (e.g., clinicopathology, physiological, and biochemical indicators) as well as treatment modalities may change. Therefore, adjustments to the regression models are absolutely necessary. In this paper, based on the cRMST, a robust dynamic RMST model is established by incorporating time-dependent covariates and time-varying effects of covariates, realizing the dynamic prediction of patients' life expectancies in the future *w* time at any prediction time $s \in [s_0, s_L]$. From the results of the C-index and prediction error, it can be seen that the prediction performance of the dynamic RMST model is better than that of the RMST model, which only uses the information at the start of follow-up (*s* = 0).

There are several considerations that need to be made when applying the dynamic RMST in practice. First, the natural cubic spline function is used in this paper to detect the time-varying effects, which is more flexible than the commonly used quadratic function. But this is not the only choice. The appropriate basis functional form can be selected according to the characteristics of the data in practice. Second, there are many choices for the structure of covariance matrix $\mathbf{V}_i$, and the independent structure (Klein and Andersen, 2005) is selected in this paper; that is, the pseudo-observations corresponding to the same individual are assumed to be independent of each other. But it is still necessary to identify the individual ID in the calculating process, otherwise the coefficient variance $\Sigma$ will be underestimated, which appeared in Nicolaie et al. (2013) and Yang et al. (2021). Aiming to overcome this defect, an improved algorithm is presented in this paper, and the simulation results corresponding to the old and new algorithms are shown in Web Table 5. The improved calculation method can effectively improve the coefficient coverage.

ACKNOWLEDGEMENTS

This work was supported by the National Natural Science Foundation of China [grant numbers 82173622, 81903411, 81673268] and the Guangdong Basic and Applied Basic Research Foundation [grant numbers 2022A1515011525, 2019A1515011506].

*Conflict of Interest*: None declared.

**Table 1.** Estimation accuracy of the cRMSTd, and type I error rates and powers of the test procedures (under 1,000,000 sample)

| n | Cen (%) | s,w | Scenario 1 (Type I error) | | | | | Scenario 2 | | | | | Scenario 3 | | | | | Scenario 4 | | | | |
|---|---|---|---|---|---|---|---|---|---|---|---|---|---|---|---|---|---|---|---|---|---|---|
| | | | Bias (×10$^2$) | RMSE | Rel SE | CP | Power | Rel bias | RMSE | Rel SE | CP | Power | Rel bias | RMSE | Rel SE | CP | Power | Rel bias | RMSE | Rel SE | CP | Power |
| 100 | 0 | 5,5 | 0.412 | 0.057 | 0.986 | 0.950 | 0.051 | 0.010 | 0.048 | 0.994 | 0.948 | 0.183 | -0.001 | 0.044 | 0.987 | 0.950 | 0.186 | 8.604* | 0.058 | 0.996 | 0.950 | 0.050 |
| | | 5,10 | 0.416 | 0.335 | 0.999 | 0.948 | 0.053 | 0.000 | 0.292 | 0.998 | 0.949 | 0.313 | -0.004 | 0.269 | 0.998 | 0.949 | 0.336 | -0.011 | 0.341 | 0.994 | 0.948 | 0.092 |
| | | 5,15 | 1.115 | 0.814 | 0.993 | 0.946 | 0.053 | 0.003 | 0.763 | 1.008 | 0.946 | 0.404 | -0.003 | 0.678 | 0.996 | 0.948 | 0.364 | 0.009 | 0.878 | 1.001 | 0.946 | 0.273 |
| | | 10,5 | 0.109 | 0.085 | 1.010 | 0.945 | 0.054 | 0.000 | 0.065 | 0.990 | 0.948 | 0.151 | -0.007 | 0.061 | 0.992 | 0.948 | 0.135 | -0.001 | 0.058 | 0.996 | 0.948 | 0.572 |
| | | 10,10 | 0.121 | 0.482 | 1.003 | 0.948 | 0.053 | -0.006 | 0.409 | 1.014 | 0.946 | 0.237 | 0.007 | 0.368 | 1.003 | 0.947 | 0.168 | 0.001 | 0.355 | 0.992 | 0.950 | 0.818 |
| | | 10,15 | -0.499 | 1.176 | 1.001 | 0.946 | 0.055 | -0.009 | 1.006 | 0.995 | 0.946 | 0.306 | -0.011 | 0.894 | 0.984 | 0.948 | 0.140 | 0.001 | 0.970 | 1.004 | 0.945 | 0.923 |
| | 15 | 5,5 | 0.064 | 0.063 | 1.009 | 0.946 | 0.054 | -0.011 | 0.052 | 1.003 | 0.948 | 0.173 | -0.004 | 0.048 | 1.002 | 0.944 | 0.178 | 0.594* | 0.062 | 1.004 | 0.949 | 0.052 |
| | | 5,10 | -0.659 | 0.368 | 1.004 | 0.945 | 0.055 | 0.014 | 0.314 | 0.999 | 0.947 | 0.301 | -0.009 | 0.290 | 0.999 | 0.949 | 0.305 | -0.005 | 0.370 | 1.004 | 0.948 | 0.090 |
| | | 5,15 | 0.181 | 0.896 | 0.991 | 0.950 | 0.050 | -0.008 | 0.807 | 0.993 | 0.950 | 0.367 | -0.010 | 0.753 | 1.007 | 0.946 | 0.342 | 0.006 | 0.916 | 0.986 | 0.950 | 0.250 |
| | | 10,5 | 0.257 | 0.095 | 1.002 | 0.947 | 0.053 | 0.000 | 0.072 | 0.984 | 0.949 | 0.133 | -0.002 | 0.068 | 0.992 | 0.949 | 0.135 | -0.006 | 0.066 | 1.000 | 0.946 | 0.499 |
| | | 10,10 | 0.887 | 0.556 | 1.004 | 0.943 | 0.057 | -0.005 | 0.445 | 0.997 | 0.947 | 0.215 | 0.008 | 0.411 | 0.992 | 0.947 | 0.145 | 0.000 | 0.418 | 1.010 | 0.945 | 0.770 |
| | | 10,15 | 0.820 | 1.395 | 1.004 | 0.948 | 0.052 | 0.007 | 1.155 | 0.995 | 0.948 | 0.286 | -0.020 | 1.057 | 0.997 | 0.948 | 0.132 | 0.003 | 1.101 | 1.004 | 0.945 | 0.888 |
| | 30 | 5,5 | -0.055 | 0.067 | 0.992 | 0.949 | 0.052 | 0.009 | 0.055 | 0.997 | 0.949 | 0.177 | -0.004 | 0.052 | 1.005 | 0.943 | 0.167 | -6.884* | 0.065 | 0.996 | 0.948 | 0.052 |
| | | 5,10 | -0.247 | 0.402 | 0.997 | 0.950 | 0.050 | -0.008 | 0.342 | 1.000 | 0.948 | 0.265 | 0.001 | 0.312 | 0.994 | 0.949 | 0.290 | 0.023 | 0.404 | 1.009 | 0.946 | 0.092 |
| | | 5,15 | 0.783 | 1.005 | 0.988 | 0.948 | 0.051 | 0.005 | 0.897 | 0.995 | 0.950 | 0.343 | 0.001 | 0.819 | 0.997 | 0.950 | 0.318 | 0.007 | 1.028 | 0.999 | 0.946 | 0.227 |
| | | 10,5 | -0.092 | 0.109 | 0.990 | 0.947 | 0.053 | -0.013 | 0.083 | 0.988 | 0.949 | 0.119 | 0.006 | 0.081 | 1.009 | 0.941 | 0.118 | -0.008 | 0.075 | 0.999 | 0.942 | 0.450 |
| | | 10,10 | -0.494 | 0.682 | 1.011 | 0.943 | 0.056 | -0.009 | 0.530 | 1.004 | 0.941 | 0.189 | 0.001 | 0.485 | 0.996 | 0.946 | 0.128 | -0.002 | 0.480 | 1.002 | 0.947 | 0.699 |
| | | 10,15 | 0.026 | 1.703 | 0.997 | 0.950 | 0.050 | 0.006 | 1.376 | 0.994 | 0.949 | 0.236 | 0.022 | 1.272 | 0.997 | 0.948 | 0.119 | 0.004 | 1.307 | 0.999 | 0.947 | 0.833 |
| | 45 | 5,5 | -0.186 | 0.077 | 1.007 | 0.946 | 0.054 | 0.005 | 0.061 | 1.000 | 0.948 | 0.157 | 0.004 | 0.056 | 0.993 | 0.947 | 0.154 | -1.110* | 0.072 | 0.998 | 0.948 | 0.052 |
| | | 5,10 | 0.211 | 0.471 | 0.997 | 0.945 | 0.055 | 0.012 | 0.391 | 1.000 | 0.948 | 0.240 | 0.001 | 0.354 | 0.992 | 0.951 | 0.257 | -0.013 | 0.457 | 1.003 | 0.945 | 0.084 |
| | | 5,15 | 0.783 | 1.281 | 1.003 | 0.944 | 0.056 | -0.001 | 1.076 | 1.000 | 0.947 | 0.299 | 0.004 | 0.979 | 1.002 | 0.946 | 0.277 | 0.006 | 1.207 | 0.993 | 0.948 | 0.206 |
| | | 10,5 | -0.121 | 0.142 | 0.988 | 0.946 | 0.055 | 0.007 | 0.108 | 1.006 | 0.940 | 0.103 | -0.003 | 0.098 | 0.992 | 0.943 | 0.096 | -0.018 | 0.093 | 0.981 | 0.938 | 0.342 |
| | | 10,10 | -0.087 | 0.922 | 0.993 | 0.943 | 0.057 | 0.000 | 0.676 | 0.985 | 0.945 | 0.147 | -0.010 | 0.634 | 0.993 | 0.945 | 0.105 | -0.003 | 0.623 | 0.980 | 0.952 | 0.586 |
| | | 10,15 | 0.992 | 2.701 | 0.993 | 0.944 | 0.056 | 0.015 | 1.973 | 0.984 | 0.947 | 0.184 | 0.000 | 1.853 | 0.997 | 0.943 | 0.104 | -0.002 | 1.935 | 0.995 | 0.943 | 0.672 |
| 200 | 0 | 5,5 | 0.014 | 0.029 | 0.996 | 0.950 | 0.050 | -0.004 | 0.024 | 0.996 | 0.950 | 0.334 | -0.010 | 0.023 | 1.005 | 0.947 | 0.336 | -0.396* | 0.029 | 1.000 | 0.949 | 0.050 |
| | | 5,10 | -0.121 | 0.165 | 0.994 | 0.951 | 0.049 | 0.003 | 0.148 | 1.009 | 0.945 | 0.549 | -0.005 | 0.134 | 0.996 | 0.950 | 0.572 | 0.011 | 0.172 | 0.999 | 0.950 | 0.146 |
| | | 5,15 | 0.608 | 0.408 | 0.995 | 0.950 | 0.050 | 0.001 | 0.378 | 1.005 | 0.946 | 0.677 | 0.002 | 0.336 | 0.993 | 0.953 | 0.619 | 0.000 | 0.435 | 0.996 | 0.950 | 0.436 |
| | | 10,5 | 0.068 | 0.041 | 0.999 | 0.949 | 0.051 | 0.008 | 0.033 | 1.008 | 0.947 | 0.261 | 0.003 | 0.030 | 0.996 | 0.949 | 0.267 | 0.006 | 0.029 | 0.998 | 0.949 | 0.844 |
| | | 10,10 | -0.084 | 0.241 | 1.006 | 0.946 | 0.054 | 0.002 | 0.199 | 1.003 | 0.947 | 0.423 | 0.000 | 0.183 | 1.002 | 0.948 | 0.290 | 0.002 | 0.180 | 1.003 | 0.947 | 0.982 |
| | | 10,15 | 1.279 | 0.587 | 1.001 | 0.948 | 0.052 | 0.008 | 0.503 | 0.998 | 0.951 | 0.543 | -0.002 | 0.460 | 1.000 | 0.949 | 0.250 | 0.000 | 0.475 | 0.999 | 0.948 | 0.998 |
| | 15 | 5,5 | -0.145 | 0.031 | 0.994 | 0.952 | 0.048 | 0.012 | 0.025 | 0.995 | 0.950 | 0.313 | -0.002 | 0.024 | 1.001 | 0.948 | 0.327 | 0.114* | 0.031 | 1.014 | 0.948 | 0.052 |
| | | 5,10 | -0.335 | 0.186 | 1.011 | 0.944 | 0.056 | 0.002 | 0.161 | 1.012 | 0.944 | 0.514 | 0.001 | 0.149 | 1.016 | 0.943 | 0.543 | -0.006 | 0.185 | 1.006 | 0.946 | 0.137 |
| | | 5,15 | 0.099 | 0.449 | 0.994 | 0.952 | 0.048 | 0.002 | 0.415 | 1.011 | 0.945 | 0.627 | -0.008 | 0.367 | 0.996 | 0.952 | 0.576 | 0.006 | 0.456 | 0.986 | 0.954 | 0.421 |
| | | 10,5 | -0.226 | 0.047 | 0.999 | 0.948 | 0.052 | -0.024 | 0.036 | 0.994 | 0.950 | 0.222 | -0.004 | 0.034 | 0.996 | 0.949 | 0.226 | -0.001 | 0.033 | 1.001 | 0.946 | 0.809 |
| | | 10,10 | -0.309 | 0.280 | 1.011 | 0.947 | 0.053 | 0.005 | 0.223 | 0.998 | 0.950 | 0.375 | 0.009 | 0.206 | 0.998 | 0.949 | 0.247 | 0.006 | 0.206 | 1.008 | 0.946 | 0.970 |
| | | 10,15 | -0.227 | 0.677 | 0.996 | 0.948 | 0.052 | 0.008 | 0.568 | 0.990 | 0.950 | 0.495 | 0.009 | 0.521 | 0.994 | 0.951 | 0.222 | 0.002 | 0.547 | 1.006 | 0.948 | 0.995 |
| | 30 | 5,5 | 0.250 | 0.034 | 0.998 | 0.948 | 0.052 | -0.002 | 0.028 | 1.008 | 0.945 | 0.302 | -0.004 | 0.026 | 1.003 | 0.949 | 0.307 | 1.015* | 0.033 | 1.009 | 0.944 | 0.056 |
| | | 5,10 | 0.298 | 0.201 | 1.000 | 0.951 | 0.049 | 0.001 | 0.169 | 0.998 | 0.951 | 0.481 | 0.003 | 0.159 | 1.003 | 0.948 | 0.509 | 0.001 | 0.190 | 0.981 | 0.952 | 0.124 |
| | | 5,15 | -0.355 | 0.510 | 0.998 | 0.949 | 0.051 | -0.004 | 0.448 | 0.997 | 0.951 | 0.574 | 0.002 | 0.403 | 0.991 | 0.949 | 0.554 | -0.002 | 0.511 | 0.996 | 0.948 | 0.391 |
| | | 10,5 | 0.166 | 0.054 | 0.990 | 0.948 | 0.052 | -0.002 | 0.044 | 1.021 | 0.942 | 0.206 | -0.013 | 0.039 | 0.998 | 0.946 | 0.204 | -0.003 | 0.037 | 0.991 | 0.949 | 0.749 |
| | | 10,10 | -0.075 | 0.330 | 1.002 | 0.947 | 0.053 | 0.006 | 0.258 | 0.997 | 0.950 | 0.337 | -0.003 | 0.233 | 0.981 | 0.952 | 0.208 | 0.001 | 0.233 | 0.994 | 0.950 | 0.943 |
| | | 10,15 | 0.563 | 0.808 | 0.978 | 0.952 | 0.048 | -0.008 | 0.700 | 1.006 | 0.947 | 0.419 | -0.005 | 0.624 | 0.992 | 0.950 | 0.188 | 0.000 | 0.624 | 0.984 | 0.951 | 0.987 |

| n | Cen | | | | | | | | | | | | | | | | | | | | |
|---|---|---|---|---|---|---|---|---|---|---|---|---|---|---|---|---|---|---|---|---|---|
| | 45 | 5,5 | 0.125 | 0.038 | 1.003 | 0.949 | 0.051 | 0.008 | 0.031 | 1.002 | 0.949 | 0.271 | 0.010 | 0.028 | 0.990 | 0.950 | 0.279 | -0.142* | 0.036 | 0.998 | 0.948 | 0.054 |
| | | 5,10 | -0.668 | 0.235 | 1.002 | 0.948 | 0.052 | -0.007 | 0.197 | 1.007 | 0.947 | 0.420 | 0.006 | 0.181 | 1.004 | 0.947 | 0.464 | 0.010 | 0.223 | 0.996 | 0.952 | 0.118 |
| | | 5,15 | 0.757 | 0.626 | 0.996 | 0.949 | 0.051 | -0.002 | 0.540 | 1.009 | 0.944 | 0.522 | -0.004 | 0.484 | 1.000 | 0.947 | 0.480 | 0.003 | 0.612 | 1.005 | 0.947 | 0.343 |
| | | 10,5 | -0.097 | 0.072 | 0.998 | 0.949 | 0.051 | -0.005 | 0.054 | 1.009 | 0.946 | 0.173 | 0.007 | 0.048 | 0.993 | 0.946 | 0.175 | -0.001 | 0.049 | 1.009 | 0.942 | 0.641 |
| | | 10,10 | -0.589 | 0.453 | 0.997 | 0.948 | 0.053 | -0.003 | 0.343 | 1.001 | 0.949 | 0.265 | 0.012 | 0.312 | 0.992 | 0.947 | 0.184 | 0.004 | 0.318 | 1.000 | 0.945 | 0.866 |
| | | 10,15 | 0.940 | 1.259 | 0.983 | 0.951 | 0.049 | -0.002 | 0.959 | 0.989 | 0.948 | 0.318 | -0.006 | 0.874 | 0.985 | 0.951 | 0.142 | 0.001 | 0.927 | 0.994 | 0.946 | 0.932 |
| 500 | 0 | 5,5 | -0.027 | 0.012 | 0.996 | 0.953 | 0.048 | -0.002 | 0.010 | 1.004 | 0.949 | 0.665 | -0.002 | 0.009 | 1.003 | 0.948 | 0.696 | 1.202* | 0.011 | 0.993 | 0.952 | 0.049 |
| | | 5,10 | -0.156 | 0.067 | 0.999 | 0.948 | 0.052 | -0.004 | 0.058 | 0.997 | 0.951 | 0.898 | -0.002 | 0.053 | 0.993 | 0.950 | 0.914 | -0.002 | 0.069 | 0.998 | 0.950 | 0.265 |
| | | 5,15 | -0.088 | 0.164 | 0.997 | 0.951 | 0.049 | 0.003 | 0.151 | 1.002 | 0.949 | 0.966 | -0.002 | 0.134 | 0.992 | 0.950 | 0.952 | -0.001 | 0.173 | 0.993 | 0.954 | 0.819 |
| | | 10,5 | 0.261 | 0.016 | 0.991 | 0.952 | 0.048 | -0.005 | 0.013 | 0.995 | 0.946 | 0.524 | -0.001 | 0.012 | 0.996 | 0.951 | 0.575 | 0.004 | 0.012 | 1.000 | 0.952 | 0.997 |
| | | 10,10 | 0.323 | 0.095 | 1.002 | 0.950 | 0.051 | -0.001 | 0.080 | 1.006 | 0.945 | 0.785 | 0.005 | 0.074 | 1.006 | 0.949 | 0.621 | -0.003 | 0.072 | 1.003 | 0.949 | 1.000 |
| | | 10,15 | -0.357 | 0.228 | 0.990 | 0.951 | 0.050 | 0.005 | 0.201 | 0.999 | 0.950 | 0.900 | 0.001 | 0.184 | 1.003 | 0.951 | 0.501 | -0.001 | 0.189 | 0.998 | 0.947 | 1.000 |
| | 15 | 5,5 | -0.142 | 0.013 | 1.008 | 0.948 | 0.052 | 0.003 | 0.010 | 1.002 | 0.949 | 0.642 | -0.003 | 0.010 | 1.009 | 0.946 | 0.678 | 2.911* | 0.012 | 1.010 | 0.949 | 0.051 |
| | | 5,10 | 0.091 | 0.073 | 1.000 | 0.949 | 0.052 | 0.005 | 0.062 | 1.000 | 0.948 | 0.879 | 0.002 | 0.058 | 1.001 | 0.950 | 0.912 | 0.001 | 0.073 | 1.002 | 0.950 | 0.270 |
| | | 5,15 | 0.344 | 0.181 | 0.999 | 0.951 | 0.048 | 0.008 | 0.161 | 0.996 | 0.949 | 0.961 | -0.001 | 0.148 | 1.002 | 0.946 | 0.929 | -0.001 | 0.187 | 1.001 | 0.952 | 0.808 |
| | | 10,5 | 0.132 | 0.019 | 1.000 | 0.950 | 0.050 | -0.008 | 0.015 | 1.013 | 0.949 | 0.478 | -0.006 | 0.014 | 1.002 | 0.948 | 0.520 | 0.002 | 0.013 | 1.005 | 0.950 | 0.995 |
| | | 10,10 | -0.283 | 0.112 | 1.011 | 0.946 | 0.054 | 0.002 | 0.090 | 1.008 | 0.949 | 0.766 | -0.002 | 0.082 | 0.995 | 0.946 | 0.569 | 0.001 | 0.082 | 1.008 | 0.949 | 1.000 |
| | | 10,15 | -0.399 | 0.276 | 1.005 | 0.950 | 0.051 | 0.002 | 0.230 | 0.999 | 0.952 | 0.862 | 0.001 | 0.209 | 0.994 | 0.951 | 0.467 | 0.002 | 0.209 | 0.985 | 0.952 | 1.000 |
| | 30 | 5,5 | 0.044 | 0.014 | 0.999 | 0.952 | 0.049 | 0.009 | 0.011 | 0.996 | 0.951 | 0.631 | 0.004 | 0.010 | 1.011 | 0.949 | 0.651 | -0.557* | 0.013 | 1.002 | 0.948 | 0.052 |
| | | 5,10 | 0.236 | 0.079 | 0.996 | 0.952 | 0.049 | 0.003 | 0.067 | 0.992 | 0.951 | 0.861 | -0.003 | 0.063 | 0.999 | 0.953 | 0.895 | 0.003 | 0.080 | 1.009 | 0.947 | 0.260 |
| | | 5,15 | 0.633 | 0.203 | 0.996 | 0.950 | 0.050 | 0.000 | 0.179 | 1.000 | 0.949 | 0.936 | 0.000 | 0.166 | 1.008 | 0.949 | 0.901 | 0.006 | 0.207 | 1.004 | 0.945 | 0.777 |
| | | 10,5 | -0.178 | 0.022 | 0.993 | 0.951 | 0.049 | 0.004 | 0.017 | 0.992 | 0.949 | 0.449 | -0.004 | 0.016 | 1.009 | 0.944 | 0.460 | 0.002 | 0.015 | 1.006 | 0.949 | 0.988 |
| | | 10,10 | 0.055 | 0.128 | 0.993 | 0.952 | 0.048 | 0.001 | 0.105 | 1.007 | 0.948 | 0.697 | 0.009 | 0.097 | 1.004 | 0.948 | 0.494 | -0.003 | 0.093 | 0.998 | 0.948 | 1.000 |
| | | 10,15 | 0.170 | 0.333 | 0.997 | 0.948 | 0.052 | 0.001 | 0.277 | 1.006 | 0.945 | 0.815 | 0.008 | 0.247 | 0.989 | 0.952 | 0.403 | 0.001 | 0.258 | 1.002 | 0.949 | 1.000 |
| | 45 | 5,5 | -0.141 | 0.015 | 0.999 | 0.952 | 0.048 | 0.005 | 0.012 | 0.991 | 0.952 | 0.582 | 0.003 | 0.011 | 0.993 | 0.952 | 0.616 | 0.177* | 0.015 | 1.002 | 0.950 | 0.051 |
| | | 5,10 | -0.077 | 0.094 | 1.001 | 0.949 | 0.051 | 0.003 | 0.077 | 1.000 | 0.951 | 0.817 | 0.001 | 0.072 | 1.003 | 0.949 | 0.840 | 0.005 | 0.090 | 1.002 | 0.949 | 0.234 |
| | | 5,15 | 0.647 | 0.249 | 0.997 | 0.953 | 0.047 | 0.002 | 0.215 | 1.007 | 0.949 | 0.888 | -0.004 | 0.189 | 0.989 | 0.952 | 0.860 | -0.004 | 0.238 | 0.994 | 0.953 | 0.700 |
| | | 10,5 | -0.138 | 0.028 | 0.993 | 0.950 | 0.050 | 0.001 | 0.021 | 0.999 | 0.950 | 0.358 | -0.009 | 0.019 | 0.999 | 0.948 | 0.374 | -0.003 | 0.019 | 1.001 | 0.947 | 0.961 |
| | | 10,10 | -0.092 | 0.178 | 0.993 | 0.950 | 0.050 | -0.004 | 0.134 | 0.994 | 0.948 | 0.569 | -0.006 | 0.124 | 0.995 | 0.951 | 0.378 | 0.000 | 0.124 | 0.994 | 0.951 | 0.998 |
| | | 10,15 | 0.222 | 0.493 | 0.985 | 0.952 | 0.048 | 0.002 | 0.376 | 0.989 | 0.949 | 0.658 | -0.003 | 0.357 | 1.007 | 0.945 | 0.280 | 0.001 | 0.366 | 0.997 | 0.948 | 1.000 |

Note: *n* is the sample size of each group, Cen is the corresponding censoring rate.

*: The true cRMST difference is approximately equal to 0, resulting in excessive relative bias.

Table 2. Estimation accuracy of the regression coefficients in the dynamic RMST model (under 100,000 sample)

| Cen (%) | Var | Linear mixed-effect model | | | | | | | | | | Quadratic mixed-effect model | | | | | | | | | |
|---|---|---|---|---|---|---|---|---|---|---|---|---|---|---|---|---|---|---|---|---|---|
| | | Bias ($\times 10^2$) | | Rel bias ($\times 10^2$) | | RMSE | | Rel SE | | CP | | Bias ($\times 10^2$) | | Rel bias ($\times 10^2$) | | RMSE | | Rel SE | | CP | |
| | | N=500 | N=1000 | N=500 | N=1000 | N=500 | N=1000 | N=500 | N=1000 | N=500 | N=1000 | N=500 | N=1000 | N=500 | N=1000 | N=500 | N=1000 | N=500 | N=1000 | N=500 | N=1000 |
| 0 | (Int) | 0.359 | -0.181 | 0.077 | -0.039 | 0.161 | 0.113 | 1.010 | 0.991 | 0.943 | 0.949 | 0.280 | 0.009 | 0.069 | 0.002 | 0.084 | 0.059 | 1.009 | 0.996 | 0.944 | 0.949 |
| | $X_1$ | -0.041 | 0.007 | 0.081 | -0.014 | 0.106 | 0.074 | 1.014 | 0.993 | 0.946 | 0.951 | -0.133 | -0.029 | 0.438 | 0.096 | 0.100 | 0.071 | 0.990 | 1.003 | 0.951 | 0.950 |
| | $X_1: s1^*$ | -0.469 | -0.621 | -3.913 | -5.180 | 0.298 | 0.214 | 0.975 | 1.005 | 0.953 | 0.948 | 0.501 | 0.096 | 1.357 | 0.262 | 0.224 | 0.158 | 0.992 | 0.992 | 0.952 | 0.953 |
| | $X_1: s2$ | -0.129 | -0.201 | -0.615 | -0.959 | 0.344 | 0.239 | 1.006 | 0.967 | 0.947 | 0.952 | 0.301 | 0.067 | 0.911 | 0.204 | 0.255 | 0.181 | 0.999 | 1.007 | 0.948 | 0.950 |
| | $X_1: s3$ | -0.891 | -0.134 | -2.610 | -0.394 | 0.337 | 0.239 | 0.986 | 1.000 | 0.946 | 0.945 | 0.285 | -0.412 | 0.874 | -1.263 | 0.272 | 0.189 | 1.017 | 0.981 | 0.945 | 0.953 |
| | $X_1: s4$ | -0.350 | -0.211 | -1.285 | -0.775 | 0.278 | 0.195 | 1.004 | 0.989 | 0.947 | 0.952 | -0.207 | -0.115 | -0.463 | -0.258 | 0.237 | 0.168 | 0.991 | 0.992 | 0.953 | 0.954 |
| | $X_1: s5$ | 0.012 | -0.569 | 0.027 | -1.298 | 0.283 | 0.196 | 1.037 | 0.998 | 0.940 | 0.948 | 0.018 | -0.073 | 0.068 | -0.270 | 0.244 | 0.173 | 1.003 | 1.009 | 0.950 | 0.952 |
| | $X_2$ | -0.014 | 0.126 | -0.030 | 0.270 | 0.065 | 0.046 | 1.000 | 1.001 | 0.947 | 0.949 | -0.059 | 0.038 | -0.191 | 0.124 | 0.067 | 0.047 | 1.002 | 1.003 | 0.948 | 0.949 |
| | $X_2: s1$ | 0.258 | 0.237 | -2.300 | -2.114 | 0.177 | 0.124 | 1.019 | 0.992 | 0.944 | 0.950 | 0.141 | 0.015 | -0.357 | -0.038 | 0.133 | 0.093 | 1.007 | 0.983 | 0.947 | 0.949 |
| | $X_2: s2$ | 0.272 | 0.069 | -1.462 | -0.372 | 0.187 | 0.131 | 1.007 | 0.986 | 0.947 | 0.949 | 0.213 | 0.190 | -0.632 | -0.561 | 0.142 | 0.100 | 1.030 | 1.023 | 0.945 | 0.945 |
| | $X_2: s3$ | 0.130 | 0.074 | -0.391 | -0.221 | 0.184 | 0.126 | 1.041 | 0.982 | 0.940 | 0.952 | -0.047 | -0.155 | 0.140 | 0.459 | 0.143 | 0.102 | 0.990 | 1.008 | 0.947 | 0.949 |
| | $X_2: s4$ | 0.155 | 0.052 | -0.612 | -0.206 | 0.161 | 0.113 | 1.018 | 1.000 | 0.945 | 0.951 | 0.289 | 0.056 | -0.586 | -0.144 | 0.146 | 0.104 | 1.007 | 1.014 | 0.946 | 0.947 |
| | $X_2: s5$ | -0.099 | -0.056 | 0.234 | 0.132 | 0.153 | 0.107 | 1.020 | 0.995 | 0.946 | 0.949 | -0.069 | -0.040 | 0.250 | 0.147 | 0.133 | 0.093 | 1.025 | 1.004 | 0.943 | 0.948 |
| | $Y(s)$ | -0.068 | 0.028 | 0.116 | -0.048 | 0.045 | 0.031 | 1.000 | 0.988 | 0.945 | 0.950 | 0.054 | 0.002 | -0.171 | -0.007 | 0.048 | 0.034 | 1.003 | 0.992 | 0.949 | 0.950 |
| | $Y(s): s1$ | -0.241 | -0.106 | 3.336 | 0.146 | 0.104 | 0.073 | 1.001 | 0.995 | 0.949 | 0.953 | -0.189 | -0.067 | 0.649 | 0.230 | 0.074 | 0.052 | 1.006 | 0.987 | 0.948 | 0.948 |
| | $Y(s): s2$ | -0.363 | -0.159 | -5.857 | -2.565 | 0.103 | 0.073 | 1.004 | 1.008 | 0.951 | 0.946 | -0.304 | -0.152 | 2.718 | 1.356 | 0.068 | 0.048 | 1.032 | 1.010 | 0.945 | 0.949 |
| | $Y(s): s3$ | -0.234 | -0.191 | -0.824 | -0.673 | 0.098 | 0.068 | 1.013 | 0.979 | 0.945 | 0.950 | -0.368 | -0.145 | -3.038 | -1.194 | 0.063 | 0.044 | 1.022 | 1.020 | 0.948 | 0.948 |
| | $Y(s): s4$ | -0.207 | -0.308 | -3.349 | -4.979 | 0.100 | 0.070 | 1.020 | 0.999 | 0.946 | 0.952 | -0.298 | -0.099 | 1.391 | 0.463 | 0.090 | 0.064 | 1.001 | 0.995 | 0.945 | 0.947 |
| | $Y(s): s5$ | -0.183 | -0.089 | -0.405 | -0.197 | 0.087 | 0.060 | 1.055 | 1.010 | 0.938 | 0.946 | -0.325 | -0.117 | -1.042 | -0.375 | 0.057 | 0.040 | 1.012 | 1.014 | 0.947 | 0.947 |
| | $s1$ | -0.148 | -0.131 | 0.161 | 0.143 | 0.358 | 0.250 | 1.017 | 0.988 | 0.944 | 0.950 | -0.550 | -0.164 | 0.489 | 0.146 | 0.214 | 0.153 | 0.997 | 1.014 | 0.947 | 0.945 |
| | $s2$ | -0.710 | 0.121 | 0.669 | -0.114 | 0.389 | 0.271 | 1.013 | 0.983 | 0.949 | 0.950 | -0.932 | -0.629 | 0.749 | 0.505 | 0.278 | 0.199 | 0.990 | 1.015 | 0.948 | 0.948 |
| | $s3$ | -0.683 | -0.363 | 0.831 | 0.442 | 0.426 | 0.291 | 1.052 | 0.981 | 0.940 | 0.948 | -1.094 | -0.111 | 1.250 | 0.127 | 0.324 | 0.231 | 1.008 | 1.018 | 0.946 | 0.945 |
| | $s4$ | -0.760 | -0.008 | 0.580 | 0.006 | 0.385 | 0.270 | 1.017 | 0.998 | 0.946 | 0.949 | -1.179 | -0.374 | 0.743 | 0.236 | 0.268 | 0.187 | 1.022 | 0.986 | 0.945 | 0.951 |
| | $s5$ | -0.459 | -0.065 | 0.925 | 0.131 | 0.385 | 0.270 | 1.031 | 1.008 | 0.938 | 0.945 | -0.654 | -0.230 | 1.523 | 0.535 | 0.307 | 0.216 | 1.021 | 0.998 | 0.942 | 0.949 |
| 15 | (Int) | 0.242 | 0.265 | 0.052 | 0.057 | 0.162 | 0.116 | 0.980 | 0.999 | 0.950 | 0.950 | 0.185 | 0.175 | 0.045 | 0.043 | 0.086 | 0.060 | 1.016 | 1.004 | 0.944 | 0.947 |
| | $X_1$ | 0.005 | -0.177 | -0.010 | 0.355 | 0.108 | 0.077 | 1.000 | 1.011 | 0.949 | 0.949 | -0.010 | -0.028 | 0.033 | 0.093 | 0.104 | 0.072 | 1.036 | 0.983 | 0.945 | 0.950 |
| | $X_1: s1$ | -0.517 | -0.246 | -4.316 | -2.053 | 0.317 | 0.225 | 0.994 | 1.001 | 0.954 | 0.948 | 0.101 | 0.086 | 0.274 | 0.234 | 0.233 | 0.163 | 0.996 | 0.976 | 0.948 | 0.953 |
| | $X_1: s2$ | -0.744 | -0.347 | -3.557 | -1.660 | 0.372 | 0.263 | 1.008 | 1.007 | 0.946 | 0.946 | 0.090 | -0.226 | 0.272 | -0.686 | 0.269 | 0.192 | 0.991 | 1.011 | 0.950 | 0.950 |
| | $X_1: s3$ | -0.600 | -0.238 | -1.757 | -0.698 | 0.384 | 0.267 | 1.040 | 1.011 | 0.938 | 0.945 | 0.341 | 0.051 | 1.046 | 0.157 | 0.296 | 0.207 | 1.026 | 1.006 | 0.947 | 0.950 |
| | $X_1: s4$ | -1.255 | -0.133 | -4.602 | -0.486 | 0.308 | 0.214 | 1.034 | 1.005 | 0.945 | 0.949 | -0.213 | -0.109 | -0.477 | -0.244 | 0.261 | 0.182 | 1.048 | 1.019 | 0.944 | 0.946 |
| | $X_1: s5$ | -0.294 | -0.115 | -0.671 | -0.263 | 0.319 | 0.222 | 1.044 | 1.011 | 0.938 | 0.944 | -0.022 | 0.093 | -0.081 | 0.344 | 0.275 | 0.190 | 1.058 | 1.009 | 0.941 | 0.949 |
| | $X_2$ | -0.009 | 0.016 | -0.020 | 0.033 | 0.066 | 0.047 | 0.988 | 0.994 | 0.949 | 0.951 | -0.072 | -0.094 | -0.235 | -0.307 | 0.069 | 0.048 | 1.038 | 1.016 | 0.941 | 0.948 |
| | $X_2: s1$ | 0.183 | 0.239 | -1.630 | -2.130 | 0.185 | 0.129 | 0.996 | 0.970 | 0.947 | 0.952 | 0.065 | -0.105 | -0.163 | 0.265 | 0.140 | 0.097 | 1.036 | 0.998 | 0.947 | 0.949 |
| | $X_2: s2$ | 0.261 | 0.105 | -1.401 | -0.561 | 0.199 | 0.143 | 0.978 | 1.103 | 0.950 | 0.948 | 0.448 | 0.178 | -1.326 | -0.527 | 0.151 | 0.103 | 1.047 | 0.974 | 0.944 | 0.952 |
| | $X_2: s3$ | 0.290 | 0.015 | -0.871 | -0.045 | 0.204 | 0.141 | 1.049 | 0.997 | 0.944 | 0.948 | -0.056 | -0.048 | 0.165 | 0.142 | 0.158 | 0.110 | 1.035 | 1.009 | 0.943 | 0.948 |
| | $X_2: s4$ | 0.674 | 0.233 | -2.659 | -0.921 | 0.176 | 0.122 | 1.042 | 1.010 | 0.943 | 0.948 | 0.541 | 0.295 | -1.096 | -0.598 | 0.156 | 0.109 | 1.038 | 1.013 | 0.943 | 0.949 |
| | $X_2: s5$ | 0.480 | 0.168 | -1.137 | -0.398 | 0.174 | 0.118 | 1.049 | 0.968 | 0.945 | 0.950 | -0.026 | 0.009 | 0.094 | -0.034 | 0.146 | 0.102 | 1.037 | 1.022 | 0.943 | 0.947 |
| | $Y(s)$ | -0.059 | -0.037 | 0.100 | 0.062 | 0.046 | 0.032 | 1.003 | 1.011 | 0.946 | 0.950 | -0.012 | -0.018 | 0.038 | 0.057 | 0.049 | 0.034 | 1.029 | 1.001 | 0.945 | 0.948 |

| | Var | | | | | | | | | | | | | | | | | | | | |
|---|---|---|---|---|---|---|---|---|---|---|---|---|---|---|---|---|---|---|---|---|---|
| | Y(s): s1 | -0.323 | 0.068 | 4.462 | -0.938 | 0.110 | 0.078 | 0.986 | 0.994 | 0.951 | 0.947 | -0.251 | -0.079 | 0.861 | 0.273 | 0.077 | 0.054 | 1.016 | 0.999 | 0.947 | 0.952 |
| | Y(s): s2 | -0.415 | -0.285 | -3.698 | -4.598 | 0.112 | 0.079 | 1.001 | 0.997 | 0.947 | 0.949 | -0.367 | -0.131 | 3.278 | 1.173 | 0.071 | 0.050 | 1.025 | 0.986 | 0.945 | 0.948 |
| | Y(s): s3 | -0.293 | -0.006 | -1.031 | -0.021 | 0.109 | 0.077 | 1.001 | 1.015 | 0.943 | 0.946 | -0.460 | -0.149 | -3.792 | -1.231 | 0.066 | 0.047 | 1.016 | 1.003 | 0.946 | 0.948 |
| | Y(s): s4 | -0.359 | -0.183 | -5.806 | -2.964 | 0.107 | 0.076 | 1.024 | 1.029 | 0.947 | 0.946 | -0.383 | -0.068 | 1.785 | 0.319 | 0.093 | 0.066 | 1.015 | 1.008 | 0.948 | 0.948 |
| | Y(s): s5 | -0.353 | -0.071 | -0.783 | -0.158 | 0.097 | 0.067 | 1.051 | 1.023 | 0.941 | 0.945 | -0.397 | -0.145 | -1.273 | -0.463 | 0.061 | 0.043 | 1.042 | 1.014 | 0.943 | 0.948 |
| | s1 | 0.065 | -0.521 | -0.071 | 0.569 | 0.368 | 0.261 | 0.971 | 0.978 | 0.950 | 0.950 | -0.448 | 0.094 | 0.398 | -0.083 | 0.226 | 0.159 | 1.029 | 1.017 | 0.943 | 0.946 |
| | s2 | -0.076 | -0.301 | 0.071 | 0.284 | 0.410 | 0.296 | 0.974 | 1.016 | 0.950 | 0.946 | -1.180 | -0.356 | 0.948 | 0.285 | 0.299 | 0.208 | 1.021 | 0.980 | 0.945 | 0.952 |
| | s3 | -0.727 | -0.267 | 0.884 | 0.324 | 0.466 | 0.321 | 1.042 | 0.983 | 0.938 | 0.948 | -1.568 | -0.336 | 1.791 | 0.383 | 0.352 | 0.250 | 1.004 | 1.015 | 0.946 | 0.947 |
| | s4 | -1.178 | -0.446 | 0.900 | 0.340 | 0.421 | 0.294 | 1.038 | 1.013 | 0.942 | 0.947 | -2.043 | -0.661 | 1.289 | 0.417 | 0.291 | 0.208 | 1.012 | 1.028 | 0.945 | 0.944 |
| | s5 | -1.485 | -0.600 | 2.993 | 1.209 | 0.436 | 0.303 | 1.056 | 1.014 | 0.934 | 0.946 | -1.474 | -0.415 | 3.430 | 0.966 | 0.345 | 0.243 | 1.048 | 1.041 | 0.938 | 0.940 |
| 30 | (Int) | -0.016 | 0.006 | -0.003 | 0.001 | 0.170 | 0.119 | 1.023 | 0.997 | 0.943 | 0.949 | 0.059 | 0.133 | 0.015 | 0.033 | 0.087 | 0.061 | 1.003 | 0.980 | 0.948 | 0.949 |
| | $X_1$ | -0.072 | 0.001 | 0.145 | -0.002 | 0.112 | 0.078 | 1.032 | 1.008 | 0.944 | 0.952 | 0.134 | 0.137 | -0.442 | -0.452 | 0.104 | 0.074 | 0.998 | 1.007 | 0.947 | 0.946 |
| | $X_1$: s1 | 0.142 | -0.103 | 1.183 | -0.856 | 0.341 | 0.242 | 1.024 | 1.034 | 0.946 | 0.948 | 0.141 | 0.230 | 0.381 | 0.623 | 0.248 | 0.172 | 1.026 | 0.982 | 0.947 | 0.951 |
| | $X_1$: s2 | -1.114 | 0.126 | -5.325 | 0.601 | 0.414 | 0.286 | 1.030 | 0.991 | 0.942 | 0.950 | -1.123 | -0.691 | -3.405 | -2.095 | 0.294 | 0.206 | 1.024 | 1.003 | 0.946 | 0.948 |
| | $X_1$: s3 | -0.825 | 0.302 | -2.417 | 0.886 | 0.444 | 0.310 | 1.048 | 1.034 | 0.937 | 0.944 | -0.484 | 0.409 | -1.482 | 1.254 | 0.328 | 0.229 | 1.028 | 1.001 | 0.946 | 0.950 |
| | $X_1$: s4 | -0.839 | -0.281 | -3.076 | -1.030 | 0.353 | 0.244 | 1.037 | 0.998 | 0.940 | 0.949 | -0.793 | -0.318 | -1.777 | -0.713 | 0.282 | 0.200 | 1.014 | 1.017 | 0.948 | 0.947 |
| | $X_1$: s5 | -0.892 | -0.140 | -2.034 | -0.318 | 0.382 | 0.267 | 1.044 | 1.031 | 0.931 | 0.940 | 0.008 | 0.108 | 0.031 | 0.397 | 0.307 | 0.214 | 1.027 | 0.998 | 0.943 | 0.947 |
| | $X_2$ | -0.004 | 0.040 | -0.009 | 0.086 | 0.068 | 0.048 | 1.017 | 1.003 | 0.946 | 0.946 | -0.110 | -0.162 | -0.361 | -0.529 | 0.069 | 0.048 | 1.013 | 0.997 | 0.947 | 0.947 |
| | $X_2$: s1 | 0.189 | 0.199 | -1.681 | -1.772 | 0.195 | 0.139 | 0.983 | 1.000 | 0.948 | 0.948 | -0.119 | -0.237 | 0.299 | 0.599 | 0.143 | 0.103 | 0.994 | 1.020 | 0.946 | 0.944 |
| | $X_2$: s2 | 0.495 | 0.363 | -2.657 | -1.952 | 0.227 | 0.155 | 1.046 | 0.985 | 0.941 | 0.953 | 0.459 | 0.406 | -1.358 | -1.203 | 0.158 | 0.112 | 1.003 | 1.002 | 0.949 | 0.950 |
| | $X_2$: s3 | 0.246 | -0.438 | -0.740 | 1.315 | 0.231 | 0.161 | 1.007 | 1.001 | 0.943 | 0.947 | -0.254 | -0.182 | 0.755 | 0.540 | 0.174 | 0.121 | 1.041 | 1.011 | 0.941 | 0.948 |
| | $X_2$: s4 | 0.597 | 0.030 | -2.357 | -0.120 | 0.199 | 0.137 | 1.046 | 1.005 | 0.945 | 0.946 | 0.533 | 0.528 | -1.080 | -1.069 | 0.167 | 0.118 | 1.023 | 1.021 | 0.945 | 0.946 |
| | $X_2$: s5 | 0.664 | -0.116 | -1.572 | 0.274 | 0.209 | 0.141 | 1.071 | 0.999 | 0.939 | 0.947 | 0.123 | -0.001 | -0.445 | 0.005 | 0.165 | 0.115 | 1.042 | 1.029 | 0.941 | 0.948 |
| | Y(s) | 0.048 | 0.020 | -0.081 | -0.034 | 0.047 | 0.033 | 1.026 | 0.987 | 0.944 | 0.952 | -0.069 | -0.082 | 0.219 | 0.258 | 0.049 | 0.035 | 0.990 | 0.981 | 0.950 | 0.951 |
| | Y(s): s1 | -0.115 | -0.003 | 1.590 | 0.040 | 0.118 | 0.083 | 0.999 | 0.998 | 0.951 | 0.949 | -0.131 | -0.107 | 0.449 | 0.367 | 0.080 | 0.056 | 0.982 | 0.973 | 0.950 | 0.952 |
| | Y(s): s2 | -0.328 | -0.388 | -5.296 | -6.251 | 0.125 | 0.087 | 1.018 | 1.031 | 0.943 | 0.954 | -0.249 | -0.034 | 2.224 | 0.302 | 0.074 | 0.053 | 0.990 | 0.995 | 0.945 | 0.950 |
| | Y(s): s3 | -0.319 | -0.099 | -1.124 | -0.348 | 0.129 | 0.090 | 1.064 | 1.000 | 0.935 | 0.942 | -0.474 | -0.188 | -3.908 | -1.550 | 0.072 | 0.050 | 1.048 | 1.006 | 0.943 | 0.947 |
| | Y(s): s4 | -0.831 | -0.359 | 13.426 | -5.796 | 0.120 | 0.083 | 1.050 | 1.008 | 0.941 | 0.950 | -0.323 | -0.057 | 1.505 | 0.264 | 0.096 | 0.068 | 1.008 | 0.992 | 0.949 | 0.951 |
| | Y(s): s5 | -0.262 | -0.146 | -0.580 | -0.323 | 0.115 | 0.079 | 1.077 | 1.007 | 0.933 | 0.943 | -0.466 | -0.228 | -1.493 | -0.730 | 0.066 | 0.045 | 1.051 | 0.991 | 0.941 | 0.949 |
| | s1 | -0.215 | -0.403 | 0.235 | 0.440 | 0.395 | 0.279 | 0.992 | 0.995 | 0.949 | 0.951 | 0.041 | 0.063 | -0.036 | -0.056 | 0.235 | 0.166 | 1.015 | 1.009 | 0.946 | 0.948 |
| | s2 | -0.018 | -0.428 | 0.017 | 0.404 | 0.463 | 0.317 | 1.043 | 0.973 | 0.939 | 0.951 | -0.335 | -0.190 | 0.269 | 0.152 | 0.320 | 0.226 | 1.015 | 1.008 | 0.944 | 0.948 |
| | s3 | -0.853 | 0.504 | 1.037 | -0.613 | 0.528 | 0.370 | 1.022 | 1.009 | 0.938 | 0.943 | -0.976 | -0.330 | 1.115 | 0.377 | 0.397 | 0.275 | 1.043 | 1.000 | 0.940 | 0.946 |
| | s4 | -0.493 | 0.342 | 0.377 | -0.261 | 0.473 | 0.330 | 1.028 | 1.001 | 0.940 | 0.945 | -1.657 | -1.157 | 1.045 | 0.730 | 0.330 | 0.233 | 1.045 | 1.034 | 0.940 | 0.943 |
| | s5 | -1.648 | -0.054 | 3.322 | 0.109 | 0.519 | 0.356 | 1.065 | 1.006 | 0.926 | 0.939 | -1.714 | -1.034 | 3.988 | 2.405 | 0.394 | 0.276 | 1.057 | 1.032 | 0.936 | 0.942 |

Abbreviations: Cen, censoring rate; Var, variable; Int, intercept.

*: The interaction terms between these two variables, in which $s1$–$s5$ represent the first to fifth spline basis of the prediction time s.

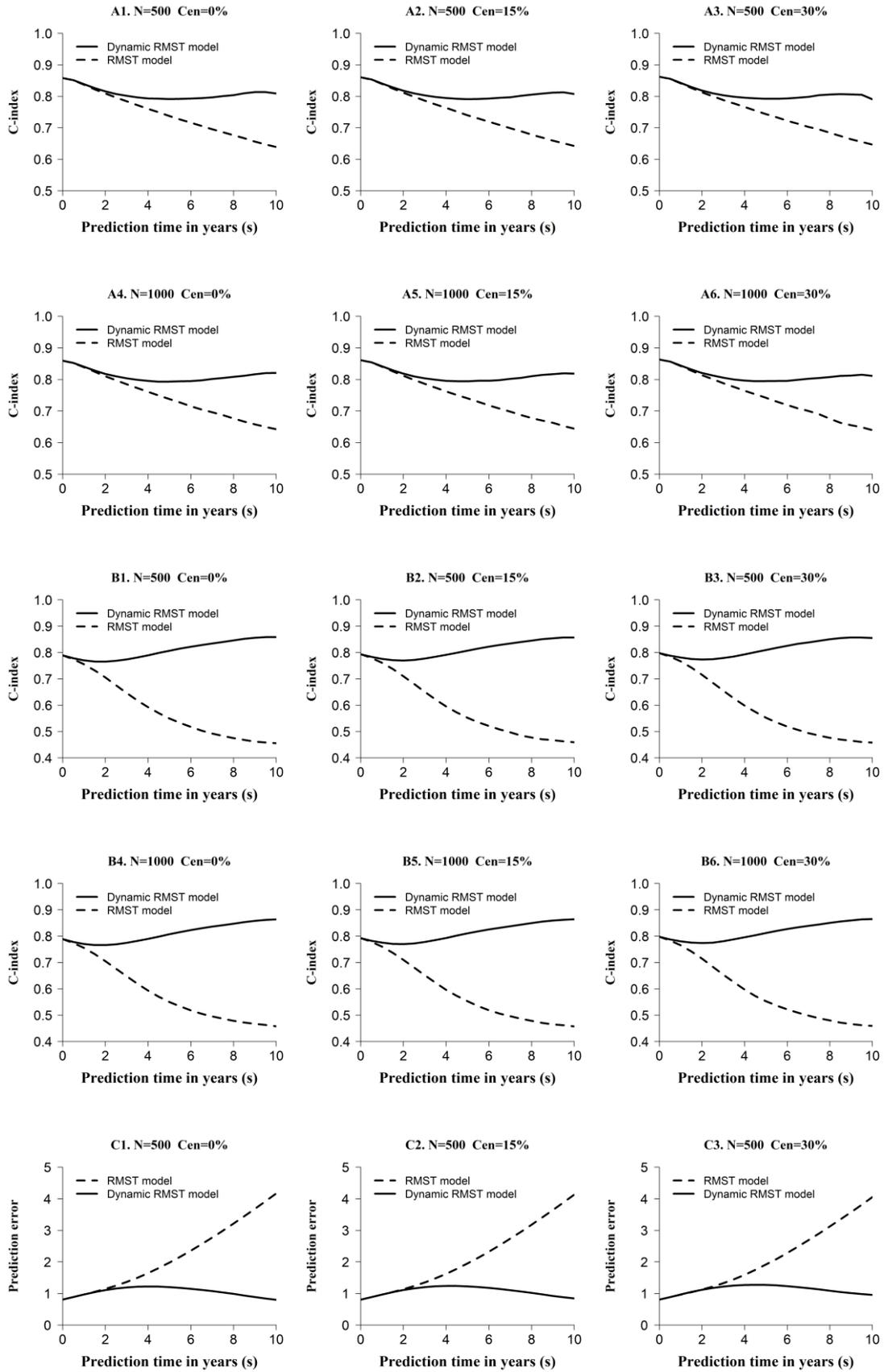

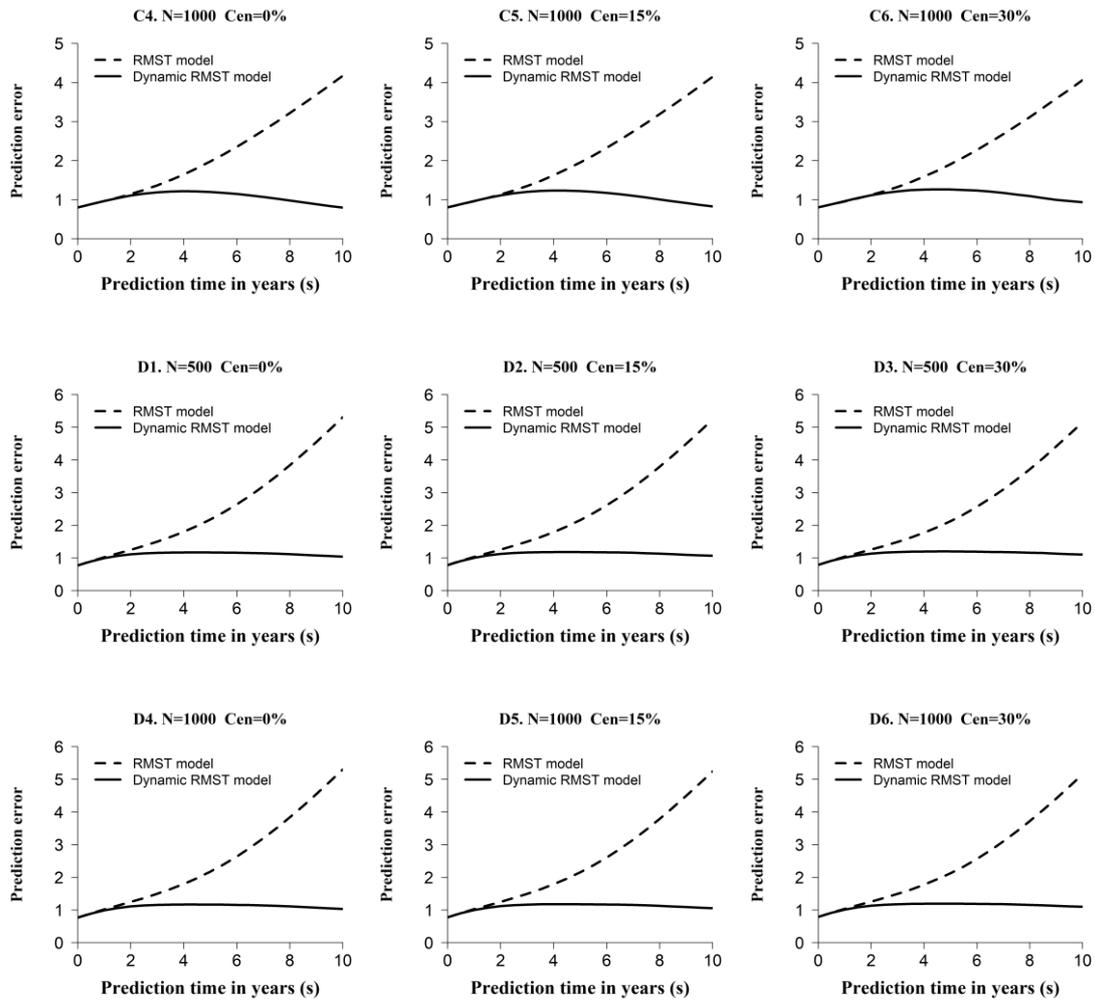

**Figure 1.** Simulation results corresponding to different sample sizes of the training set and the censoring rate for landmark time-specific C-indexes and prediction errors

A1–A6 represent the C-indexes of the linear mixed-effects model, and C1–C6 represent the corresponding prediction errors. B1–B6 represent the C-indexes of the quadratic mixed-effects model, and D1–D6 represent the corresponding prediction errors. A higher C-index indicates a better performing model; a lower prediction error indicates a better performing model.

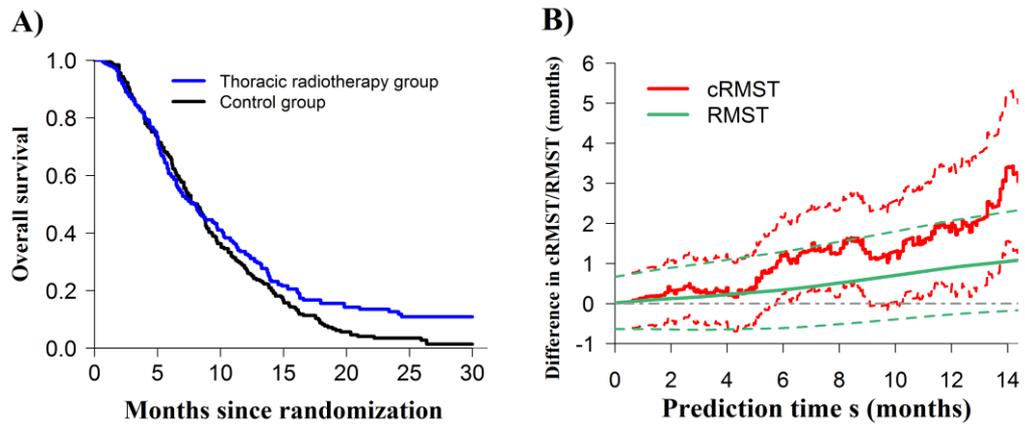

**Figure 2.** Results based on the example trial data set.

(A) displays the Kaplan–Meier survival curves by treatment group. (B) displays the differences in 12-month cRMST or ($s$+12)-month RMST with 95% confidence intervals.

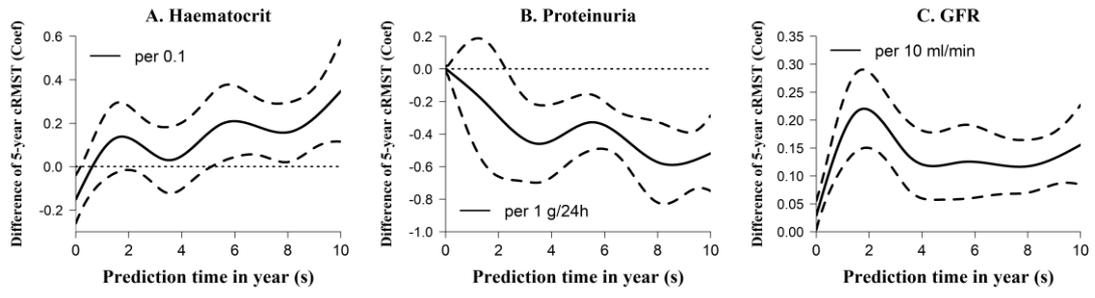

**Figure 3.** Differences in 5-year cRMST (dynamic coefficients $\beta(s)$) with 95% confidence intervals in the dynamic RMST model ($w$ = 5 years)

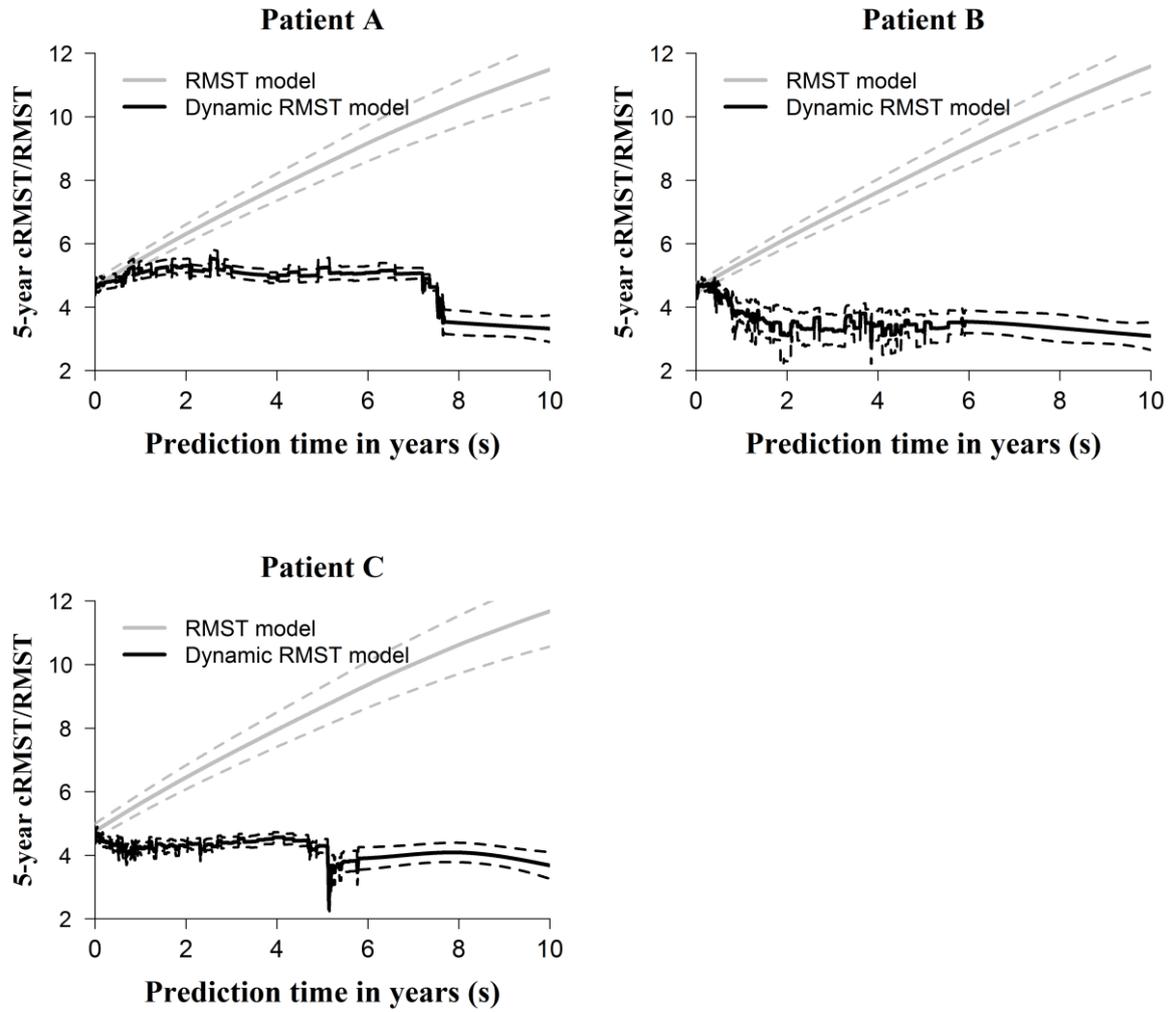

**Figure 4.** Individual predictions with the dynamic RMST model ($w = 5$ years) and the RMST model ($\tau = s+5$ years)

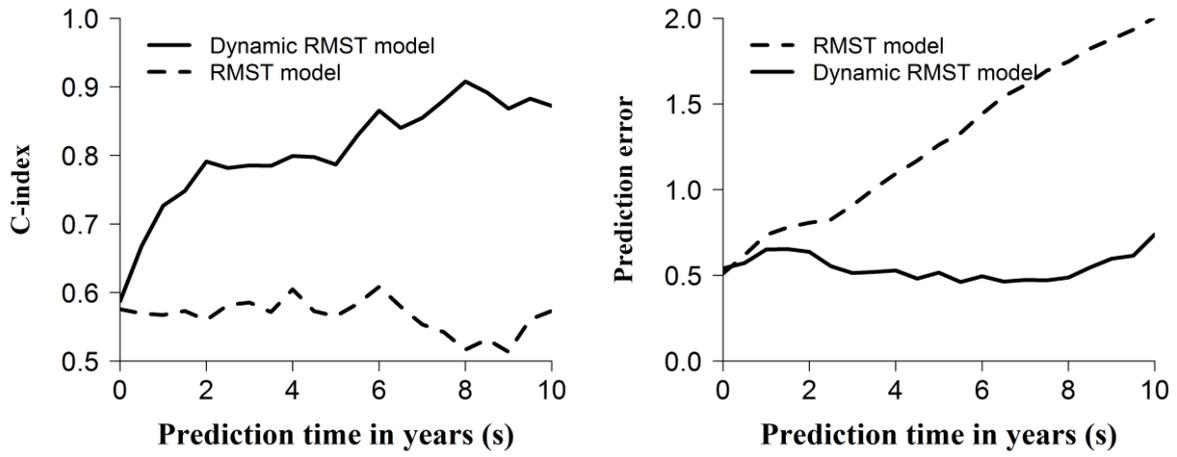

**Figure 5.** Landmark time-specific C-indexes and prediction errors

# Supporting Informatin for "Analysis of dynamic restricted mean survival time based on pseudo-observations" by Zijing Yang, Chengfeng Zhang, Yawen Hou and Zheng Chen

Zijing Yang[1,2], Chengfeng Zhang[2], Yawen Hou[3*], Zheng Chen[2**]

[1]Stomatological Hospital, Southern Medical University and Guangdong Provincial Stomatological Hospital, 366 Jiangnan Avenue South, Guangzhou, 510260, China

[2]Department of Biostatistics, School of Public Health (Guangdong Provincial Key Laboratory of Tropical Disease Research), Southern Medical University, Guangzhou, 510515, China

[3]Department of Statistics, School of Economics, Jinan University, Guangzhou, 510632, China

## A  Consistency proof of two approaches for estimation of the cRMST

Suppose that the events occur at $D$ distinct times $t_1 < t_2 < ... < t_D$, and that at time $t_k$ there are $d_k$ events. Let $Y_k$ be the number of individuals who are at risk at time $t_k$. The standard estimator of the survival function proposed by Kaplan–Meier is:

$$\hat{S}(t) = \begin{cases} 1 & , t < t_1 \\ \prod_{t_k \leq t}[1-\frac{d_k}{Y_k}] & , t_1 \leq t \end{cases}.$$

Defining $t_{m-1} \leq s < t_m < ... < t_n \leq s+w < t_{n+1}$, the cRMST can be calculated by

$$\begin{aligned}
\hat{\mu}_{KM}(s,w) &= E(\min(T-s,w)\,|\,T>s) = \frac{\int_s^{s+w} \hat{S}(t)dt}{\hat{S}(s)} \\
&= \frac{\hat{S}(t_{m-1})(t_m - s) + \sum_{i=m}^{n-1} \hat{S}(t_i)(t_{i+1}-t_i) + \hat{S}(t_n)(s+w-t_n)}{\hat{S}(s)} \\
&= (t_m - s) + \sum_{i=m}^{n-1} \frac{\hat{S}(t_i)}{\hat{S}(s)}(t_{i+1}-t_i) + \frac{\hat{S}(t_n)}{\hat{S}(s)}(s+w-t_n) \\
&= (t_m - s) + \sum_{i=m}^{n-1} \frac{\prod_{t_k \leq t_i}[1-\frac{d_k}{Y_k}]}{\prod_{t_k \leq s}[1-\frac{d_k}{Y_k}]}(t_{i+1}-t_i) + \frac{\prod_{t_k \leq t_n}[1-\frac{d_k}{Y_k}]}{\prod_{t_k \leq s}[1-\frac{d_k}{Y_k}]}(s+w-t_n) \\
&= (t_m - s) + \sum_{i=m}^{n-1} \prod_{s < t_k \leq t_i}[1-\frac{d_k}{Y_k}](t_{i+1}-t_i) + \prod_{s < t_k \leq t_n}[1-\frac{d_k}{Y_k}](s+w-t_n)
\end{aligned}.$$

The survival function corresponding to the patients at risk at $s$ is denoted as $S_s(t)$, which can also be estimated by the Kaplan–Meier method:

$$\hat{S}_s(t) = \begin{cases} 1 & , t < t_m \\ \prod_{s < t_k \leq t}[1-\frac{d_k}{Y_k}] & , t \geq t_m \end{cases},$$

then $\hat{\mu}_{KM}(s,w) = (t_m - s) + \sum_{i=m}^{n-1} \hat{S}_s(t_i)(t_{i+1}-t_i) + \hat{S}_s(t_n)(s+w-t_n) = \int_s^{s+w} \hat{S}_s(t)dt$.

**Web Table 1.** The results of the hypothesis test ($w$=12 months)

| Prediction time $s$ (month) | cRMSTd (95% CI) | $Z$ | $P$ |
|---|---|---|---|
| 0 | 0.016 (-0.635, 0.667) | 0.048 | 0.962 |
| 2 | 0.324 (-0.429, 1.078) | 0.844 | 0.398 |
| 4 | 0.395 (-0.447, 1.236) | 0.919 | 0.358 |
| 6 | 1.192 (0.239, 2.145) | 2.450 | 0.014 |
| 8 | 1.252 (0.165, 2.340) | 2.257 | 0.024 |
| 10 | 1.304 (0.040, 2.569) | 2.022 | 0.043 |
| 12 | 2.002 (0.490, 3.514) | 2.595 | 0.009 |
| 14 | 3.405 (1.558, 5.253) | 3.613 | <0.001 |

**Web Table 2.** The results of the dynamic RMST model (*w*=5 years)

| Variable | Coefficient | SE | *P* value |
|---|---|---|---|
| (Intercept) | 5.537 | 0.217 | <0.001 |
| Age (per 10 years) | 0.097 | 0.025 | <0.001 |
| Weight (per 10 Kg) | -0.125 | 0.027 | <0.001 |
| Sex (ref: female) | | | |
| Male | -0.112 | 0.059 | 0.057 |
| Hematocrit (per 0.1) | -0.150 | 0.057 | 0.008 |
| Hematocrit: *s1*\* | 0.045 | 0.146 | 0.760 |
| Hematocrit: *s2* | 0.475 | 0.169 | 0.005 |
| Hematocrit: *s3* | 0.102 | 0.139 | 0.464 |
| Hematocrit: *s4* | 0.761 | 0.166 | <0.001 |
| Hematocrit: *s5* | 0.255 | 0.140 | 0.069 |
| Proteinuria (per 1 g/24h) | 0.001 | 0.008 | 0.888 |
| Proteinuria: *s1* | -0.597 | 0.194 | 0.002 |
| Proteinuria: *s2* | -0.167 | 0.163 | 0.305 |
| Proteinuria: *s3* | -0.624 | 0.241 | 0.010 |
| Proteinuria: *s4* | -0.746 | 0.270 | 0.006 |
| Proteinuria: *s5* | -0.396 | 0.214 | 0.064 |
| GFR (per 10 ml/min) | 0.030 | 0.013 | 0.026 |
| GFR: *s1* | 0.059 | 0.050 | 0.238 |
| GFR: *s2* | 0.113 | 0.055 | 0.041 |
| GFR: *s3* | 0.000 | 0.041 | 0.993 |
| GFR: *s4* | 0.335 | 0.060 | <0.001 |
| GFR: *s5* | -0.030 | 0.047 | 0.522 |
| *s1* | -0.307 | 0.567 | 0.588 |
| *s2* | -2.287 | 0.649 | <0.001 |
| *s3* | -0.028 | 0.572 | 0.961 |
| *s4* | -4.387 | 0.623 | <0.001 |
| *s5* | -0.647 | 0.590 | 0.273 |

\*: The interaction terms between these two variables, in which *s*1–*s*5 represent the first to fifth spline basis of the prediction time *s*.

**Web Table 3.** The results of the RMST model ($\tau = 15$ years)

| Variable | Coef | 95% CI | Z | P |
|---|---|---|---|---|
| **(Intercept)** | 15.253 | (12.115, 18.390) | 9.528 | <0.001 |
| **Age** (per 10 years) | 0.394 | (-0.008, 0.795) | 1.922 | 0.055 |
| **Weight** (per 10 Kg) | -0.318 | (-0.774, -0.137) | -1.369 | 0.171 |
| **Haematocrit** (per 0.1) | -0.649 | (-0.006, -1.292) | -1.977 | 0.048 |
| **Proteinuria** (per 1 g/24h) | 0.044 | (-0.017, 0.105) | 1.411 | 0.158 |
| **GFR** (per 10 ml/min) | -0.422 | (-1.249, 0.406) | -0.998 | 0.318 |
| **Sex** (ref: female) | | | | |
| Male | -0.422 | (-1.590, 0.411) | -0.998 | 0.318 |

**Web Table 4.** The definition of example patients

| Patient | Variables | | | | | | |
|---|---|---|---|---|---|---|---|
| | Time* | Age | Weight | Haematocrit | Proteinuria | GFR | Sex |
| A | 0.00 | 25 | 65.8 | 0.26 | 3.8 | 12.82 | male |
| | 1.00 | 25 | 65.8 | 0.41 | 0.0 | 94.74 | male |
| | 6.00 | 25 | 65.8 | 0.42 | 0.0 | 83.06 | male |
| | 7.00 | 25 | 65.8 | 0.39 | 0.0 | 82.56 | male |
| | 7.59 | 25 | 65.8 | 0.39 | 1.0 | 13.63 | male |
| | 7.64 | 25 | 65.8 | 0.39 | 1.4 | 17.52 | male |
| B | 0.00 | 44 | 77.2 | 0.32 | 0.3 | 6.60 | male |
| | 0.38 | 44 | 77.2 | 0.38 | 0.0 | 85.40 | male |
| | 2.00 | 44 | 77.2 | 0.28 | 2.2 | 24.01 | male |
| | 6.00 | 44 | 77.2 | 0.34 | 1.6 | 13.87 | male |
| C | 0.00 | 18 | 59.7 | 0.21 | 0.6 | 10.76 | male |
| | 4.00 | 18 | 59.7 | 0.42 | 0.0 | 48.95 | male |
| | 5.14 | 18 | 59.7 | 0.32 | 4.2 | 11.82 | male |
| | 6.00 | 18 | 59.7 | 0.32 | 0.0 | 5.41 | male |

*: The observation time. Only the data of some representative time points are intercepted for illustration.

**Web Table 5.** Simulation results of coefficient variance estimation accuracy of dynamic RMST model using methods in old and proposed algorithm

| Cen (%) | Var | Linear mixed-effect model | | | | | | | | Quadratic mixed-effect model | | | | | | | |
|---|---|---|---|---|---|---|---|---|---|---|---|---|---|---|---|---|---|
| | | Old algorithm* | | | | Proposed algorithm | | | | Old algorithm* | | | | Proposed algorithm | | | |
| | | Rel SE | | CP | | Rel SE | | CP | | Rel SE | | CP | | Rel SE | | CP | |
| | | N=500 | N=1000 | N=500 | N=1000 | N=500 | N=1000 | N=500 | N=1000 | N=500 | N=1000 | N=500 | N=1000 | N=500 | N=1000 | N=500 | N=1000 |
| 0 | (Int) | 1.572 | 1.540 | 0.879 | 0.884 | 1.010 | 0.991 | 0.943 | 0.949 | 1.481 | 1.462 | 0.889 | 0.896 | 1.009 | 0.996 | 0.944 | 0.949 |
| | $X_1$ | 1.480 | 1.449 | 0.893 | 0.893 | 1.014 | 0.993 | 0.946 | 0.951 | 1.500 | 1.520 | 0.893 | 0.892 | 0.990 | 1.003 | 0.951 | 0.950 |
| | $X_1$: $s1$[a] | 2.017 | 2.081 | 0.826 | 0.823 | 0.975 | 1.005 | 0.953 | 0.948 | 1.830 | 1.831 | 0.850 | 0.850 | 0.992 | 0.992 | 0.952 | 0.953 |
| | $X_1$: $s2$ | 1.638 | 1.576 | 0.877 | 0.884 | 1.006 | 0.967 | 0.947 | 0.952 | 1.457 | 1.467 | 0.896 | 0.894 | 0.999 | 1.007 | 0.948 | 0.950 |
| | $X_1$: $s3$ | 1.805 | 1.830 | 0.857 | 0.854 | 0.986 | 1.000 | 0.946 | 0.945 | 1.806 | 1.745 | 0.855 | 0.859 | 1.017 | 0.981 | 0.945 | 0.953 |
| | $X_1$: $s4$ | 1.070 | 1.054 | 0.943 | 0.944 | 1.004 | 0.989 | 0.947 | 0.952 | 0.969 | 0.972 | 0.954 | 0.956 | 0.991 | 0.992 | 0.953 | 0.954 |
| | $X_1$: $s5$ | 2.037 | 1.965 | 0.825 | 0.833 | 1.037 | 0.998 | 0.940 | 0.948 | 2.035 | 2.047 | 0.828 | 0.831 | 1.003 | 1.009 | 0.950 | 0.952 |
| | $X_2$ | 1.480 | 1.479 | 0.892 | 0.892 | 1.000 | 1.001 | 0.947 | 0.949 | 1.664 | 1.664 | 0.874 | 0.869 | 1.002 | 1.003 | 0.948 | 0.949 |
| | $X_2$: $s1$ | 2.117 | 2.058 | 0.821 | 0.827 | 1.019 | 0.992 | 0.944 | 0.950 | 2.043 | 1.995 | 0.831 | 0.838 | 1.007 | 0.983 | 0.947 | 0.949 |
| | $X_2$: $s2$ | 1.593 | 1.555 | 0.878 | 0.885 | 1.007 | 0.986 | 0.947 | 0.949 | 1.523 | 1.512 | 0.888 | 0.888 | 1.030 | 1.023 | 0.945 | 0.945 |
| | $X_2$: $s3$ | 1.890 | 1.778 | 0.844 | 0.858 | 1.041 | 0.982 | 0.940 | 0.952 | 1.826 | 1.862 | 0.850 | 0.851 | 0.990 | 1.008 | 0.947 | 0.949 |
| | $X_2$: $s4$ | 1.015 | 0.995 | 0.948 | 0.951 | 1.018 | 1.000 | 0.945 | 0.951 | 1.028 | 1.035 | 0.945 | 0.947 | 1.007 | 1.014 | 0.946 | 0.947 |
| | $X_2$: $s5$ | 2.083 | 2.039 | 0.819 | 0.828 | 1.020 | 0.995 | 0.946 | 0.949 | 2.200 | 2.160 | 0.813 | 0.815 | 1.025 | 1.004 | 0.943 | 0.948 |
| | $Y(s)$ | 1.564 | 1.543 | 0.886 | 0.885 | 1.000 | 0.988 | 0.945 | 0.950 | 1.889 | 1.867 | 0.843 | 0.849 | 1.003 | 0.992 | 0.949 | 0.950 |
| | $Y(s)$: $s1$ | 2.179 | 2.160 | 0.815 | 0.815 | 1.001 | 0.995 | 0.949 | 0.953 | 2.304 | 2.258 | 0.800 | 0.806 | 1.006 | 0.987 | 0.948 | 0.948 |
| | $Y(s)$: $s2$ | 2.686 | 1.687 | 0.866 | 0.867 | 1.004 | 1.008 | 0.951 | 0.946 | 1.612 | 1.575 | 0.875 | 0.879 | 1.032 | 1.010 | 0.945 | 0.949 |
| | $Y(s)$: $s3$ | 2.027 | 1.955 | 0.829 | 0.835 | 1.013 | 0.979 | 0.945 | 0.950 | 2.357 | 2.341 | 0.793 | 0.799 | 1.022 | 1.020 | 0.948 | 0.948 |
| | $Y(s)$: $s4$ | 0.980 | 0.958 | 0.951 | 0.957 | 1.020 | 0.999 | 0.946 | 0.952 | 1.047 | 1.041 | 0.939 | 0.941 | 1.001 | 0.995 | 0.945 | 0.947 |
| | $Y(s)$: $s5$ | 2.213 | 2.111 | 0.811 | 0.820 | 1.055 | 1.010 | 0.938 | 0.946 | 2.830 | 2.824 | 0.753 | 0.752 | 1.012 | 1.014 | 0.947 | 0.947 |
| | $s1$ | 2.116 | 2.055 | 0.821 | 0.828 | 1.017 | 0.988 | 0.944 | 0.950 | 1.794 | 1.826 | 0.859 | 0.851 | 0.997 | 1.014 | 0.947 | 0.945 |
| | $s2$ | 1.627 | 1.578 | 0.873 | 0.875 | 1.013 | 0.983 | 0.949 | 0.950 | 1.545 | 1.586 | 0.885 | 0.879 | 0.990 | 1.015 | 0.948 | 0.948 |
| | $s3$ | 2.072 | 1.936 | 0.823 | 0.842 | 1.052 | 0.981 | 0.940 | 0.948 | 1.811 | 1.830 | 0.850 | 0.854 | 1.008 | 1.018 | 0.946 | 0.945 |
| | $s4$ | 1.063 | 1.043 | 0.943 | 0.945 | 1.017 | 0.998 | 0.946 | 0.949 | 1.363 | 1.321 | 0.906 | 0.916 | 1.022 | 0.986 | 0.945 | 0.951 |
| | $s5$ | 2.223 | 2.183 | 0.806 | 0.813 | 1.031 | 1.008 | 0.938 | 0.945 | 1.976 | 1.937 | 0.834 | 0.838 | 1.021 | 0.998 | 0.942 | 0.949 |
| 15 | (Int) | 1.519 | 1.548 | 0.886 | 0.887 | 0.980 | 0.999 | 0.950 | 0.950 | 1.479 | 1.462 | 0.889 | 0.890 | 1.016 | 1.004 | 0.944 | 0.947 |
| | $X_1$ | 1.449 | 1.464 | 0.898 | 0.896 | 1.000 | 1.011 | 0.949 | 0.949 | 1.558 | 1.477 | 0.882 | 0.894 | 1.036 | 0.983 | 0.945 | 0.950 |
| | $X_1$: $s1$ | 2.029 | 2.044 | 0.830 | 0.831 | 0.994 | 1.001 | 0.954 | 0.948 | 1.823 | 1.786 | 0.856 | 0.856 | 0.996 | 0.976 | 0.948 | 0.953 |
| | $X_1$: $s2$ | 1.637 | 1.637 | 0.872 | 0.874 | 1.008 | 1.007 | 0.946 | 0.946 | 1.445 | 1.475 | 0.901 | 0.896 | 0.991 | 1.011 | 0.950 | 0.950 |
| | $X_1$: $s3$ | 1.890 | 1.835 | 0.846 | 0.854 | 1.040 | 1.011 | 0.938 | 0.945 | 1.810 | 1.773 | 0.854 | 0.861 | 1.026 | 1.006 | 0.947 | 0.950 |
| | $X_1$: $s4$ | 1.184 | 1.150 | 0.917 | 0.933 | 1.034 | 1.005 | 0.945 | 0.949 | 1.089 | 1.060 | 0.942 | 0.941 | 1.048 | 1.019 | 0.944 | 0.946 |
| | $X_1$: $s5$ | 2.017 | 1.956 | 0.829 | 0.834 | 1.044 | 1.011 | 0.938 | 0.944 | 2.118 | 2.023 | 0.818 | 0.830 | 1.058 | 1.009 | 0.941 | 0.949 |
| | $X_2$ | 1.452 | 1.459 | 0.829 | 0.895 | 0.988 | 0.994 | 0.949 | 0.951 | 1.711 | 1.674 | 0.865 | 0.870 | 1.038 | 1.016 | 0.941 | 0.948 |
| | $X_2$: $s1$ | 2.055 | 2.000 | 0.825 | 0.834 | 0.996 | 0.970 | 0.947 | 0.952 | 2.080 | 2.006 | 0.830 | 0.833 | 1.036 | 0.998 | 0.947 | 0.949 |
| | $X_2$: $s2$ | 1.549 | 1.598 | 0.884 | 0.879 | 0.978 | 1.103 | 0.950 | 0.948 | 1.544 | 1.436 | 0.886 | 0.896 | 1.047 | 0.974 | 0.944 | 0.952 |
| | $X_2$: $s3$ | 1.886 | 1.788 | 0.848 | 0.855 | 1.049 | 0.997 | 0.944 | 0.948 | 1.886 | 1.837 | 0.845 | 0.850 | 1.035 | 1.009 | 0.943 | 0.948 |
| | $X_2$: $s4$ | 1.104 | 1.068 | 0.937 | 0.942 | 1.042 | 1.010 | 0.943 | 0.948 | 1.102 | 1.073 | 0.936 | 0.941 | 1.038 | 1.013 | 0.943 | 0.949 |
| | $X_2$: $s5$ | 2.096 | 1.941 | 0.819 | 0.837 | 1.049 | 0.968 | 0.945 | 0.950 | 2.180 | 2.156 | 0.812 | 0.819 | 1.037 | 1.022 | 0.943 | 0.947 |
| | $Y(s)$ | 1.567 | 1.576 | 0.882 | 0.882 | 1.003 | 1.011 | 0.946 | 0.950 | 1.927 | 1.875 | 0.840 | 0.850 | 1.029 | 1.001 | 0.945 | 0.948 |
| | $Y(s)$: $s1$ | 2.159 | 2.174 | 0.819 | 0.815 | 0.986 | 0.994 | 0.951 | 0.947 | 2.337 | 2.293 | 0.800 | 0.805 | 1.016 | 0.999 | 0.947 | 0.952 |
| | $Y(s)$: $s2$ | 1.710 | 1.695 | 0.863 | 0.868 | 1.001 | 0.997 | 0.947 | 0.949 | 1.604 | 1.544 | 0.876 | 0.882 | 1.025 | 0.986 | 0.945 | 0.948 |
| | $Y(s)$: $s3$ | 2.011 | 2.029 | 0.826 | 0.825 | 1.001 | 1.015 | 0.943 | 0.946 | 2.312 | 2.272 | 0.798 | 0.809 | 1.016 | 1.003 | 0.946 | 0.948 |
| | $Y(s)$: $s4$ | 1.033 | 1.037 | 0.948 | 0.945 | 1.024 | 1.029 | 0.947 | 0.946 | 1.064 | 1.057 | 0.842 | 0.943 | 1.015 | 1.008 | 0.948 | 0.948 |
| | $Y(s)$: $s5$ | 2.184 | 2.122 | 0.711 | 0.820 | 1.051 | 1.023 | 0.941 | 0.945 | 2.838 | 2.756 | 0.754 | 0.761 | 1.042 | 1.014 | 0.943 | 0.948 |
| | $s1$ | 2.003 | 2.019 | 0.833 | 0.835 | 0.971 | 0.978 | 0.950 | 0.950 | 1.835 | 1.816 | 0.849 | 0.850 | 1.029 | 1.017 | 0.943 | 0.946 |
| | $s2$ | 1.561 | 1.626 | 0.883 | 0.875 | 0.974 | 1.016 | 0.950 | 0.946 | 1.591 | 1.531 | 0.877 | 0.884 | 1.021 | 0.980 | 0.945 | 0.952 |
| | $s3$ | 2.021 | 1.909 | 0.832 | 0.846 | 1.042 | 0.983 | 0.938 | 0.948 | 1.802 | 1.819 | 0.854 | 0.848 | 1.004 | 1.015 | 0.946 | 0.947 |
| | $s4$ | 1.152 | 1.127 | 0.932 | 0.936 | 1.038 | 1.013 | 0.942 | 0.947 | 1.447 | 1.472 | 0.895 | 0.893 | 1.012 | 1.028 | 0.945 | 0.944 |
| | $s5$ | 2.223 | 2.145 | 0.803 | 0.815 | 1.056 | 1.014 | 0.934 | 0.946 | 2.024 | 2.014 | 0.826 | 0.828 | 1.048 | 1.041 | 0.938 | 0.940 |
| 30 | (Int) | 1.576 | 1.536 | 0.881 | 0.886 | 1.023 | 0.997 | 0.943 | 0.949 | 1.450 | 1.416 | 0.895 | 0.904 | 1.003 | 0.980 | 0.948 | 0.949 |
| | $X_1$ | 1.481 | 1.445 | 0.888 | 0.898 | 1.032 | 1.008 | 0.944 | 0.952 | 1.489 | 1.503 | 0.894 | 0.890 | 0.998 | 1.007 | 0.947 | 0.946 |

|  | | | | | | | | | | | | | | | |
|---|---|---|---|---|---|---|---|---|---|---|---|---|---|---|---|
| $X_1$: $s1$ | 2.050 | 2.072 | 0.826 | 0.827 | 1.024 | 1.034 | 0.946 | 0.948 | 1.857 | 1.778 | 0.850 | 0.859 | 1.026 | 0.982 | 0.947 | 0.951 |
| $X_1$: $s2$ | 1.662 | 1.599 | 0.870 | 0.877 | 1.030 | 0.991 | 0.942 | 0.950 | 1.490 | 1.461 | 0.891 | 0.892 | 1.024 | 1.003 | 0.946 | 0.948 |
| $X_1$: $s3$ | 1.879 | 1.854 | 0.847 | 0.849 | 1.048 | 1.034 | 0.937 | 0.944 | 1.797 | 1.747 | 0.855 | 0.860 | 1.028 | 1.001 | 0.946 | 0.950 |
| $X_1$: $s4$ | 1.344 | 1.291 | 0.909 | 0.918 | 1.037 | 0.998 | 0.940 | 0.949 | 1.152 | 1.158 | 0.932 | 0.931 | 1.014 | 1.017 | 0.948 | 0.947 |
| $X_1$: $s5$ | 1.979 | 1.963 | 0.825 | 0.832 | 1.044 | 1.031 | 0.931 | 0.940 | 2.030 | 1.976 | 0.826 | 0.837 | 1.027 | 0.998 | 0.943 | 0.947 |
| $X_2$ | 1.483 | 1.463 | 0.892 | 0.893 | 1.017 | 1.003 | 0.946 | 0.946 | 1.656 | 1.630 | 0.875 | 0.880 | 1.013 | 0.997 | 0.947 | 0.947 |
| $X_2$: $s1$ | 2.004 | 2.034 | 0.837 | 0.828 | 0.983 | 1.000 | 0.948 | 0.948 | 1.977 | 2.027 | 0.835 | 0.832 | 0.994 | 1.020 | 0.946 | 0.944 |
| $X_2$: $s2$ | 1.648 | 1.555 | 0.872 | 0.883 | 1.046 | 0.985 | 0.941 | 0.953 | 1.475 | 1.474 | 0.892 | 0.894 | 1.003 | 1.002 | 0.949 | 0.950 |
| $X_2$: $s3$ | 1.791 | 1.772 | 0.858 | 0.858 | 1.007 | 1.001 | 0.943 | 0.947 | 1.866 | 1.811 | 0.846 | 0.857 | 1.041 | 1.011 | 0.941 | 0.948 |
| $X_2$: $s4$ | 1.240 | 1.187 | 0.925 | 0.929 | 1.046 | 1.005 | 0.945 | 0.946 | 1.154 | 1.151 | 0.930 | 0.932 | 1.023 | 1.021 | 0.945 | 0.946 |
| $X_2$: $s5$ | 2.087 | 1.956 | 0.818 | 0.834 | 1.071 | 0.999 | 0.939 | 0.947 | 2.145 | 2.123 | 0.817 | 0.819 | 1.042 | 1.029 | 0.941 | 0.948 |
| $Y(s)$ | 1.595 | 1.533 | 0.877 | 0.888 | 1.026 | 0.987 | 0.944 | 0.952 | 1.844 | 1.826 | 0.847 | 0.852 | 0.990 | 0.981 | 0.950 | 0.951 |
| $Y(s)$: $s1$ | 2.198 | 2.191 | 0.810 | 0.813 | 0.999 | 0.998 | 0.951 | 0.949 | 2.264 | 2.240 | 0.807 | 0.809 | 0.982 | 0.973 | 0.950 | 0.952 |
| $Y(s)$: $s2$ | 1.762 | 1.690 | 0.858 | 0.868 | 1.018 | 1.031 | 0.943 | 0.954 | 1.557 | 1.562 | 0.880 | 0.878 | 0.990 | 0.995 | 0.945 | 0.950 |
| $Y(s)$: $s3$ | 2.125 | 2.068 | 0.816 | 0.824 | 1.064 | 1.000 | 0.935 | 0.942 | 2.349 | 2.245 | 0.798 | 0.807 | 1.048 | 1.006 | 0.943 | 0.947 |
| $Y(s)$: $s4$ | 1.154 | 1.109 | 0.933 | 0.939 | 1.050 | 1.008 | 0.941 | 0.950 | 1.063 | 1.048 | 0.943 | 0.947 | 1.008 | 0.992 | 0.949 | 0.951 |
| $Y(s)$: $s5$ | 2.197 | 2.062 | 0.808 | 0.824 | 1.077 | 1.007 | 0.933 | 0.943 | 2.758 | 2.602 | 0.762 | 0.774 | 1.051 | 0.991 | 0.941 | 0.949 |
| $s1$ | 2.018 | 2.021 | 0.831 | 0.828 | 0.992 | 0.995 | 0.949 | 0.951 | 1.794 | 1.782 | 0.857 | 0.857 | 1.015 | 1.009 | 0.946 | 0.948 |
| $s2$ | 1.655 | 1.553 | 0.872 | 0.883 | 1.043 | 0.973 | 0.939 | 0.951 | 1.583 | 1.573 | 0.880 | 0.882 | 1.015 | 1.008 | 0.944 | 0.948 |
| $s3$ | 1.949 | 1.921 | 0.834 | 0.839 | 1.022 | 1.009 | 0.938 | 0.943 | 1.866 | 1.790 | 0.848 | 0.858 | 1.043 | 1.000 | 0.940 | 0.946 |
| $s4$ | 1.271 | 1.238 | 0.918 | 0.919 | 1.028 | 1.001 | 0.940 | 0.945 | 1.636 | 1.627 | 0.874 | 0.876 | 1.045 | 1.034 | 0.940 | 0.943 |
| $s5$ | 2.184 | 2.074 | 0.799 | 0.820 | 1.065 | 1.006 | 0.926 | 0.939 | 2.041 | 1.999 | 0.823 | 0.833 | 1.057 | 1.032 | 0.936 | 0.942 |

Abbreviations: Cen, censoring rate; Var, variable; Int, intercept.

*: Nicolaie et al. (2013) and Yang et al. (2021).

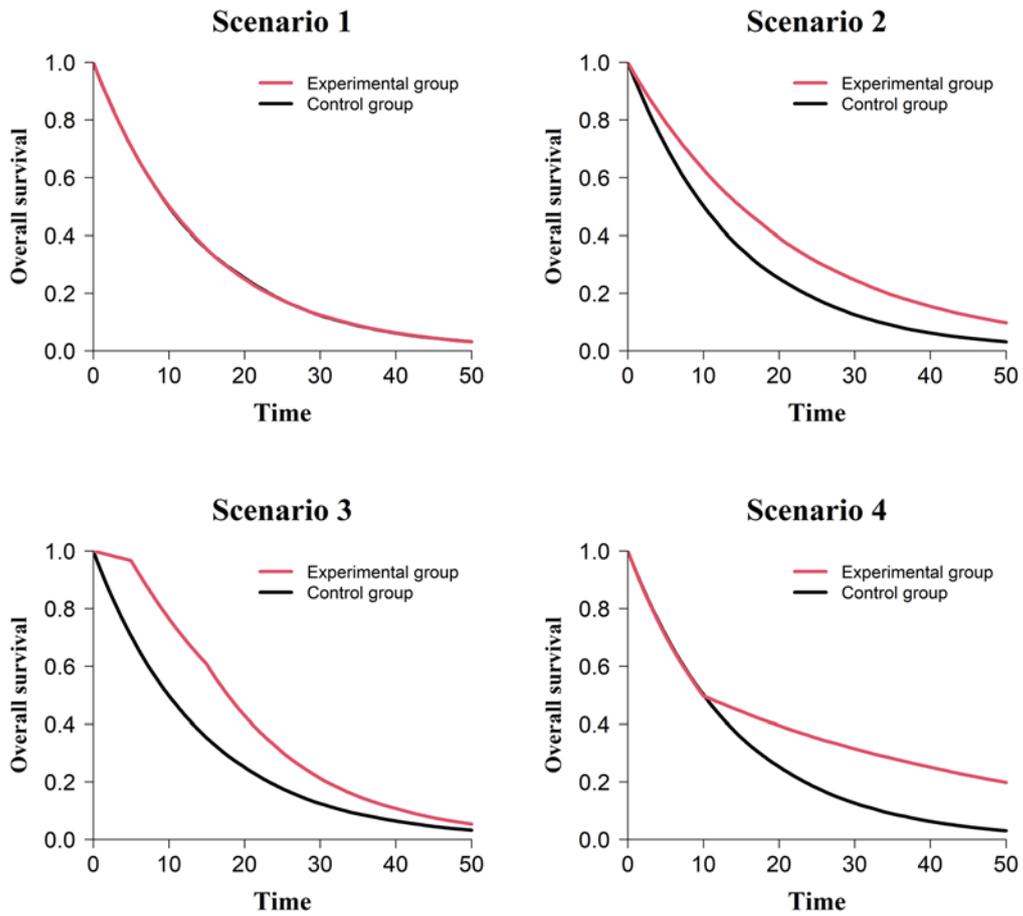

**Web Figure 1.** Four scenarios in the simulation study for the parameter estimation and hypothesis test of the cRMSTd